\documentclass[onecolumn,11pt,letterpaper]{revtex4-1}

\usepackage{graphicx}
\usepackage{epstopdf}
\usepackage{amsmath}
\usepackage{amsfonts}
\usepackage{bbm}
\usepackage{braket}

\usepackage[normalem]{ulem}

\usepackage[symbol]{footmisc}

\usepackage{color}
\usepackage{subfigure}
\usepackage[margin=1.5cm]{geometry}
\makeatletter

\newcommand{\Rmnum}[1]{\expandafter\@slowromancap\romannumeral #1@}

\makeatother

\begin{document}
\title{A Note on \emph{A Priori} Forecasting and Simplicity Bias in Time Series}

\author{Kamaludin Dingle}
\thanks{Corresponding author}
\affiliation{Centre for Applied Mathematics and Bioinformatics,
Department of Mathematics and Natural Sciences, 
Gulf University for Science and Technology, Kuwait\\
email: dingle.k@gust.edu.kw}

\author{Rafiq Kamal}

\affiliation{Centre for Applied Mathematics and Bioinformatics,
Department of Mathematics and Natural Sciences, 
Gulf University for Science and Technology, Kuwait}

\affiliation{Department of English, Gulf University for Science and Technology, Kuwait\\
email: kamal.r@gust.edu.kw}

\author{Boumediene Hamzi}
\affiliation{Department of Computing and Mathematical Sciences, \\ Caltech, Pasadena, CA, USA \\ email: boumediene.hamzi@gmail.com}

\date{\today}

\begin{abstract}
\noindent  To what extent can we forecast a time series without fitting to historical data? Can universal patterns of probability help in this task? Deep relations between pattern Kolmogorov complexity and pattern  probability have recently been used to make \emph{a priori} probability predictions in a variety of systems in physics, biology and engineering. Here we study \emph{simplicity bias} (SB) --- an exponential upper bound decay in pattern probability with increasing complexity --- in discretised time series extracted from the World Bank Open Data collection. We  predict upper bounds on the probability of discretised series patterns, without fitting to trends in the data. Thus we perform a kind of `forecasting without training data', predicting time series shape patterns \emph{a priori}, but not the actual numerical value of the series.
  Additionally we make predictions about which of two discretised series is more likely with accuracy of $\sim$80\%, much higher than a 50\% baseline rate, just by using the complexity of each series. These results point to a promising perspective on practical time series forecasting and integration with machine learning methods.\\
\vspace{0.25cm}

\end{abstract}

\maketitle

\section{Introduction}

Suppose you were challenged to forecast a time series, but you did not have any historical data from which to build or fit a model. Because essentially all time series predictions are made from fitting to past observations, this challenge is impossible to meet in full generality. Is it nonetheless possible to make some generic predictions about series patterns which might apply in a broad range of contexts? Are some patterns intrinsically more likely than others, and could these be used to forecast?  A related question has been studied (albeit in an abstract way) in a branch of computer science known as \emph{algorithmic information theory} \cite{solomonoff1960preliminary,kolmogorov1965three,chaitin1975theory,li2008introduction} (AIT). The central quantity of AIT is \emph{Kolmogorov complexity}, $K(x)$, which measures the complexity of an individual object or pattern $x$ via the amount of information required to describe or generate $x$. Within AIT, Levin's  coding theorem \cite{levin1974laws} establishes a fundamental connection between complexity and probability predictions in very general settings. Loosely, the theorem states that when  considering the chances that some output $x$ (eg a binary string, a discrete pattern, a shape) is produced via some generic computation mechanism, if $x$ is simple it is much more likely to appear than if $x$ is complex. In this manner, certain patterns are intrinsically more likely than others if generated via a computational process, and the probability of the pattern $x$ can be estimated just from the complexity of that pattern.  More formally, the coding theorem states that   $P(x)\sim 2^{-K(x)}$ where $P(x)$ is the probability that an output $x$ is generated by a (prefix optimal) universal Turing machine fed with a random binary program input. $P(x)$ is known as the \emph{algorithmic probability} of $x$, and the associated probability distribution is known as the \emph{universal distribution} \cite{solomonoff2003kolmogorov}.  Hence, under general settings, given no other information, predicting that a given simple output appears is a better bet than a more complex one. This preference for simplicity is closely connected  to Occam's razor, a fundamental principle of scientific reasoning that simpler explanations or models take preference over more complex ones \cite{sep-simplicity}. However, Occam's razor is typically used for model selection \cite{hansen2001model} rather than forecasting.

\begin{figure*}[htp]
\begin{center}
\subfigure[]{\label{fig:edge-a}\includegraphics[height=5.5cm,width=5.5cm]{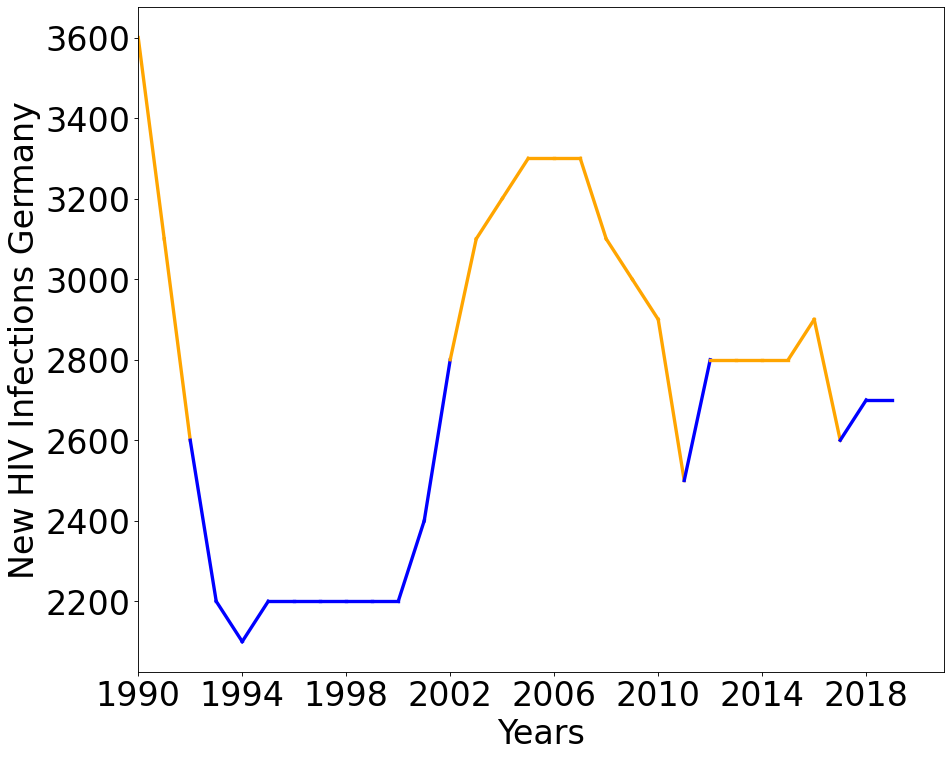}}
\subfigure[]{\label{fig:edge-a}\includegraphics[height=5.5cm,width=5.5cm]{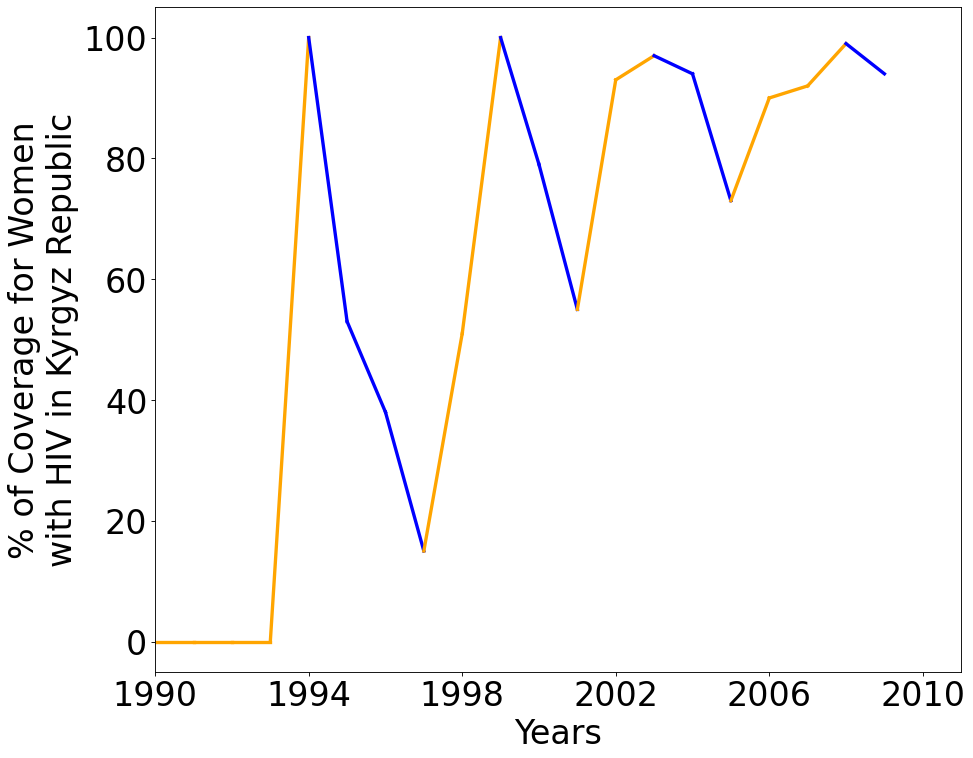}}
\subfigure[]{\label{fig:edge-a}\includegraphics[height=5.5cm,width=5.5cm]{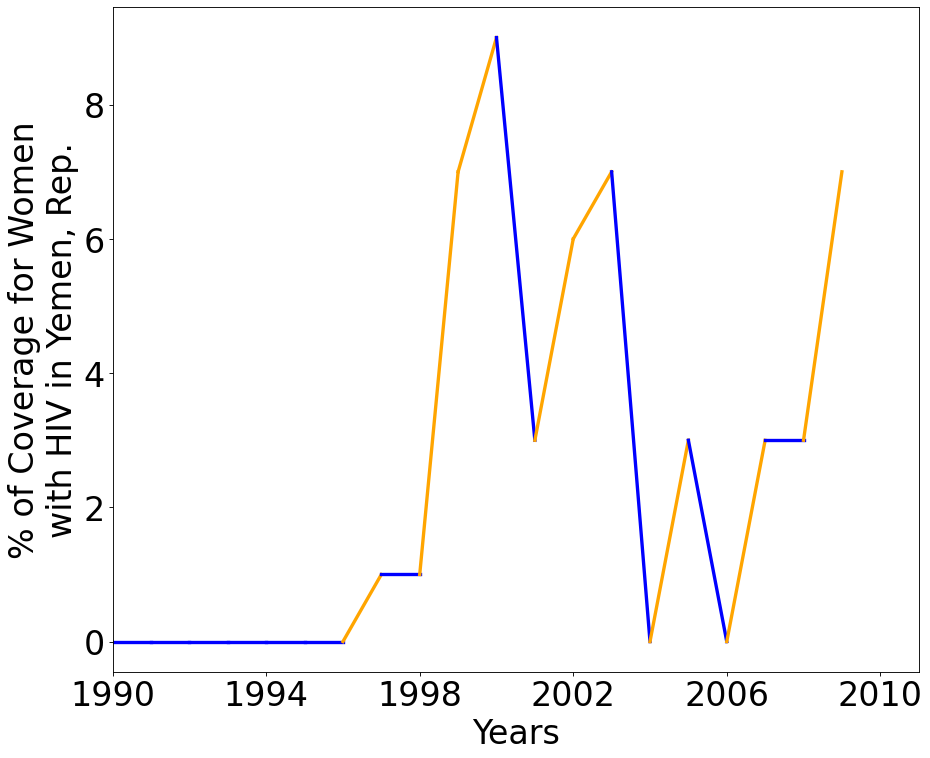}}
\end{center}
\caption{Illustrating the different binarising methods for example time series. We choose HIV rates in different countries purely for illustrative purposes. (a) High/Low (HL) method, values above the mean of the series $s(t)$ are drawn in yellow and assigned a 1 in the binary string $x(t)$, and values below in blue assigned a 0 in the binary string. (b) The Up/Down (UD) method, where regions with a positive slope are drawn in yellow assigned 1, and negative or flat drawn in yellow and assigned a 0; (c) the Up/Down method after applying a linear detrend ($Det$) is similar to UD method.}
\label{fig:blueyellow}
\end{figure*}

Algorithmic probability and  AIT results are typically difficult to apply in real-world settings due to the fact that $K(x)$ is uncomputable, the theorems assume the presence of universal Turing machines (UTMs), and the results are asymptotic and stated with accuracy to within an unknown constant. Despite these theoretical difficulties, in practice many successful applications of AIT have been made based, for example in bioinformatics \cite{cilibrasi2005clustering,ferragina2007compression}, physics \cite{avinery2019universal}, signal denoising \cite{vitanyi2013similarity}, among many other applications \cite{li2008introduction}. Mostly these applications use standard compression algorithms to approximate $K(x)$, sometimes combined with various forms of theorem approximation.  Moreover, algorithmic probability estimates have successfully been made  via a weaker form of Levin's coding theorem, applicable in real-world contexts \cite{dingle2018input}. This weaker form was applied in a range of input-output maps to make \emph{a priori} predictions regarding the probability of different shapes and patterns, such as the probability of different RNA shapes appearing on a random choice of genetic sequence, or the probability of differential equation solution profile shapes, on random choice of input parameters, and several other examples \cite{dingle2018input,dingle2020generic}. Surprisingly, it was found that probability estimates could be made directly from the complexities of the shapes themselves, without recourse to the details of the map or reference to how the shapes were generated. The authors of \cite{dingle2018input} termed this phenomenon of an inverse relation between complexity and probability \emph{simplicity bias} (SB). In a biological context, these SB results have been used to predict the frequency in nature of various biological shapes and patterns, specifically self-assembling tiles (polyominos), protein structures, natural RNA structures, and genetic network profiles \cite{johnston2022symmetry} (see also  \cite{dingle2022predicting}). Earlier, pioneering work on algorithmic probability estimates was numerically studied via small computing devices \cite{delahaye2012numerical,soler2014calculating,zenil2019coding} and sampling random computer programs \cite{legg2013approximation}. Taken together, these results suggest that algorithmic probability can be usefully applied to real-world settings to make probability predictions, in a wide range of settings. 

In this work, we investigate SB in real-world time series and test to what extent predictions can be made for real series, just based on generic universal probability assignment arguments, without recourse to fitting to training data. By assuming that the patterns observed in natural time series result from some kind of computational process, and assuming that SB predictions therefore apply, we make a form of forecasting without historical data, ie making \emph{a priori} predictions about the probability of different discretised time series patterns. Note that in this work we are not arguing that the outputs of Turing machines should be taken as the `correct' prior for time series patterns; indeed Turing machines may produce a whole array of series patterns atypical of real-world time series. Rather the premise of the work is that, based on the success of probability predictions using SB, we will perform numerical experiments to test if similar success can be achieved for time series. 

Note that the purpose of this work is not a general investigation of simplicity and complexity measures of time series \cite{tang2015complexity,torres2000relative}, nor does it argue for one complexity measure over another \cite{bialek2001complexity}. What is meant by `complexity' \cite{lloyd2001measures} and which measure is most appropriate to measure this quantity is not our focus. Rather, we are interested in applying arguments inspired by AIT --- and especially SB --- as a mathematical framework to make non-trivial training data-free probability predictions, and (estimated) Kolmogorov complexity plays a central role in this work.

\section{Results}

\begin{figure*}[htp]
\begin{center}
\subfigure[]{\label{fig:edge-a}\includegraphics[height=6.0cm,width=6.0cm]{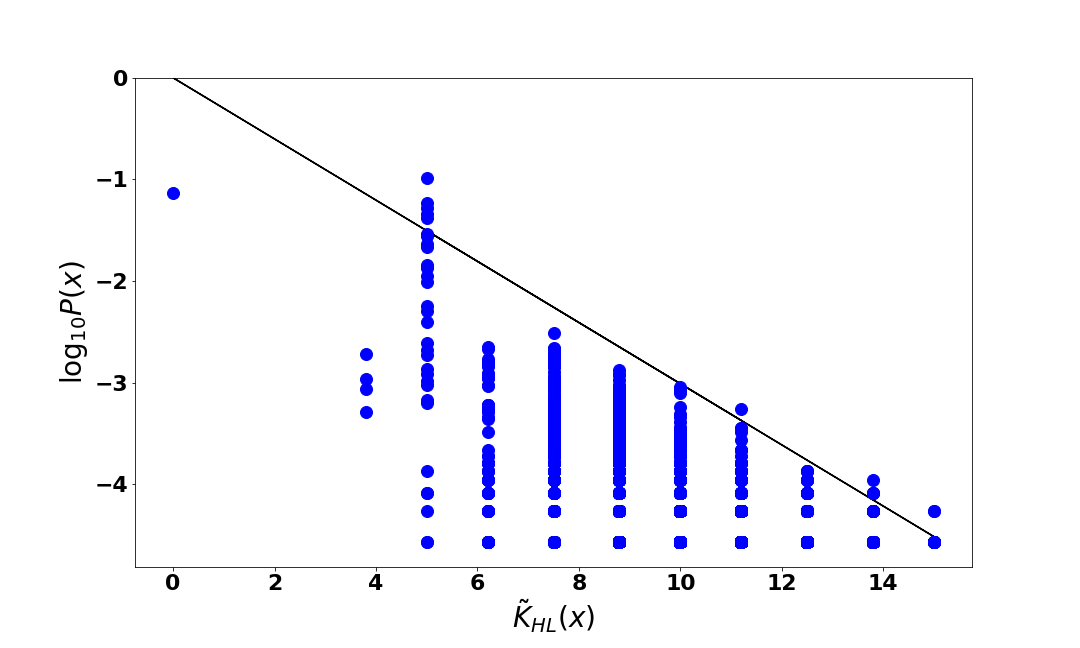}}
\subfigure[]{\label{fig:edge-a}\includegraphics[height=6.0cm,width=6.0cm]{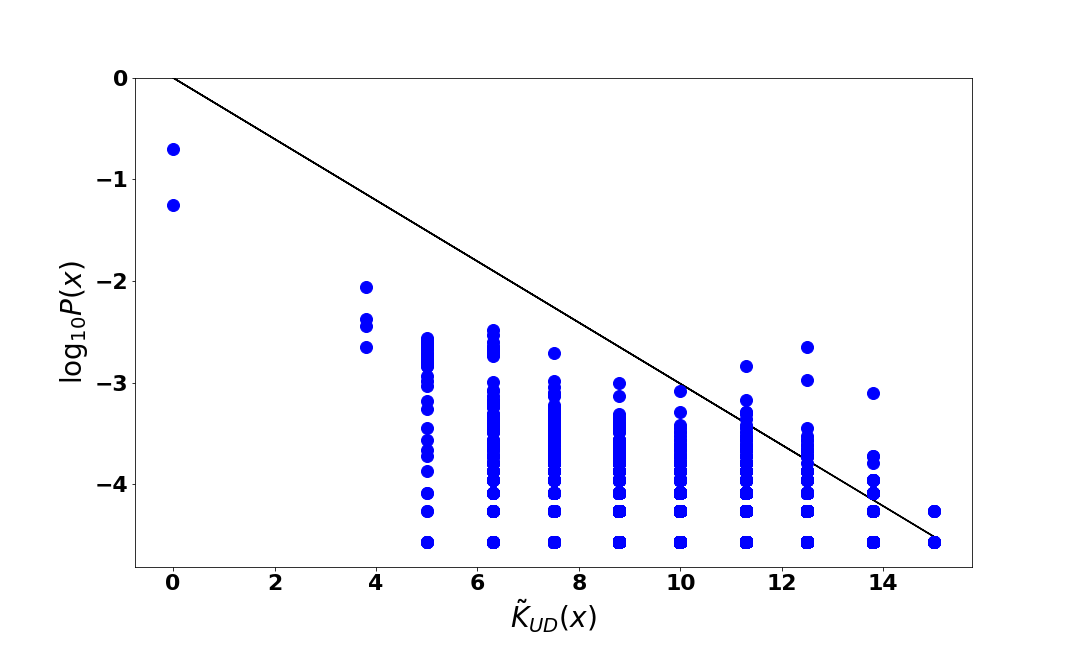}}
\subfigure[]{\label{fig:edge-a}\includegraphics[height=6.0cm,width=6.0cm]{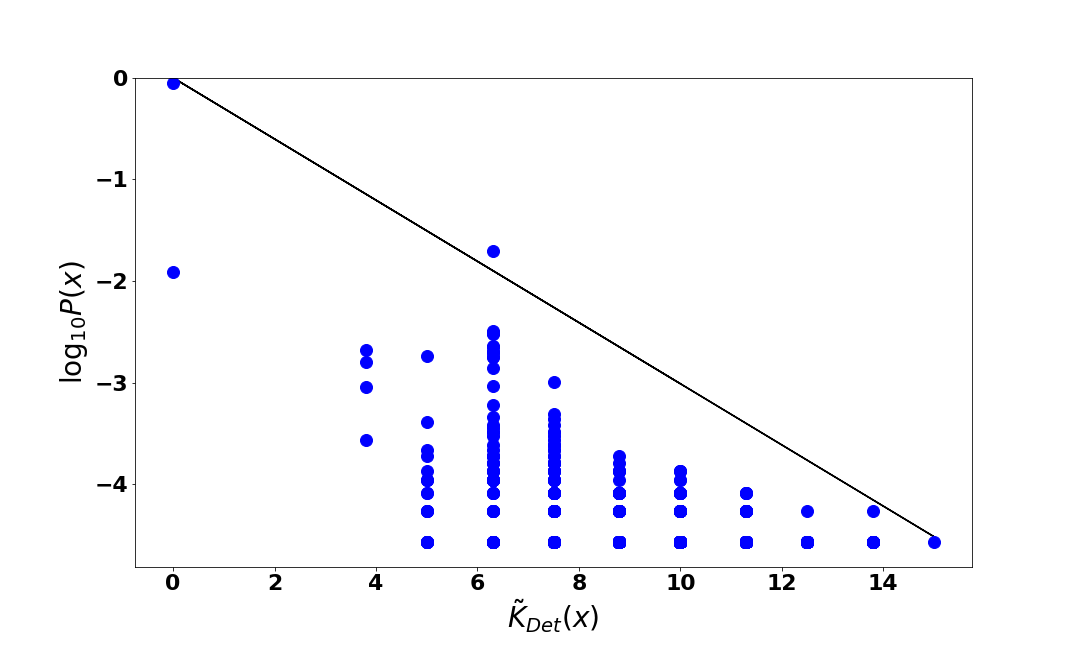}}
\subfigure[]{\label{fig:edge-a}\includegraphics[height=6.0cm,width=6.0cm]{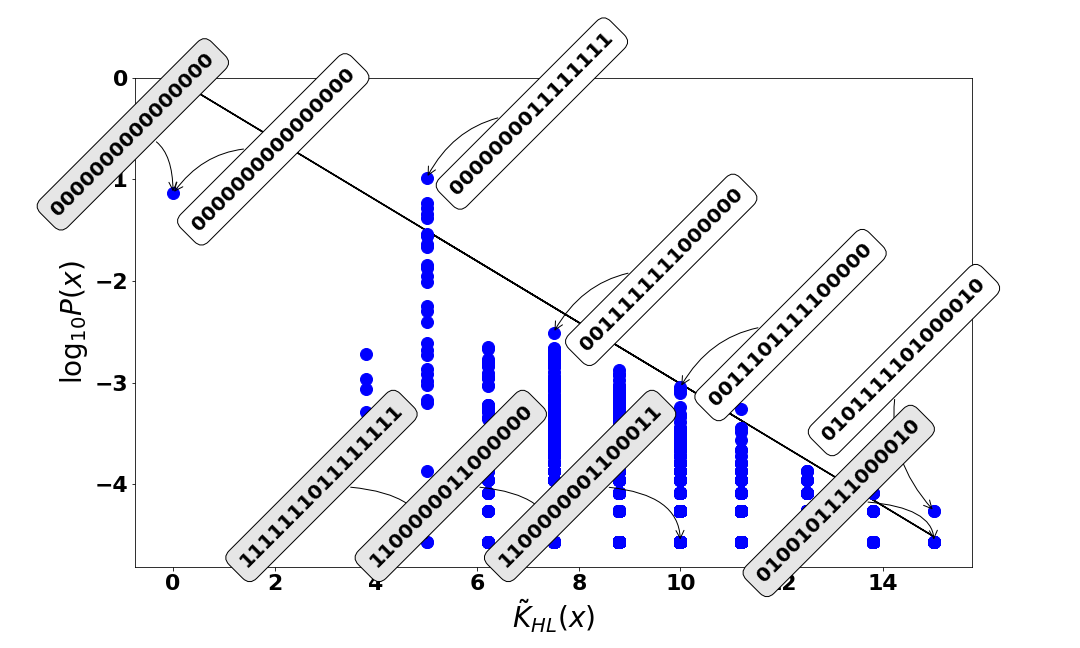}}
\subfigure[]{\label{fig:edge-a}\includegraphics[height=6.0cm,width=6.0cm]{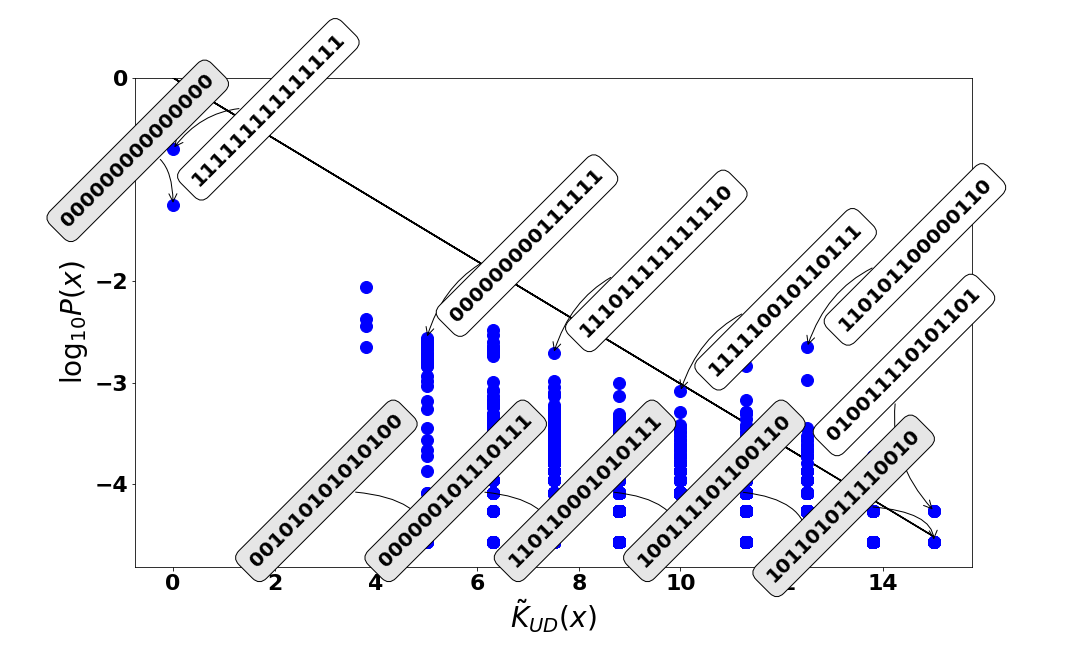}}
\subfigure[]{\label{fig:edge-a}\includegraphics[height=6.0cm,width=6.0cm]{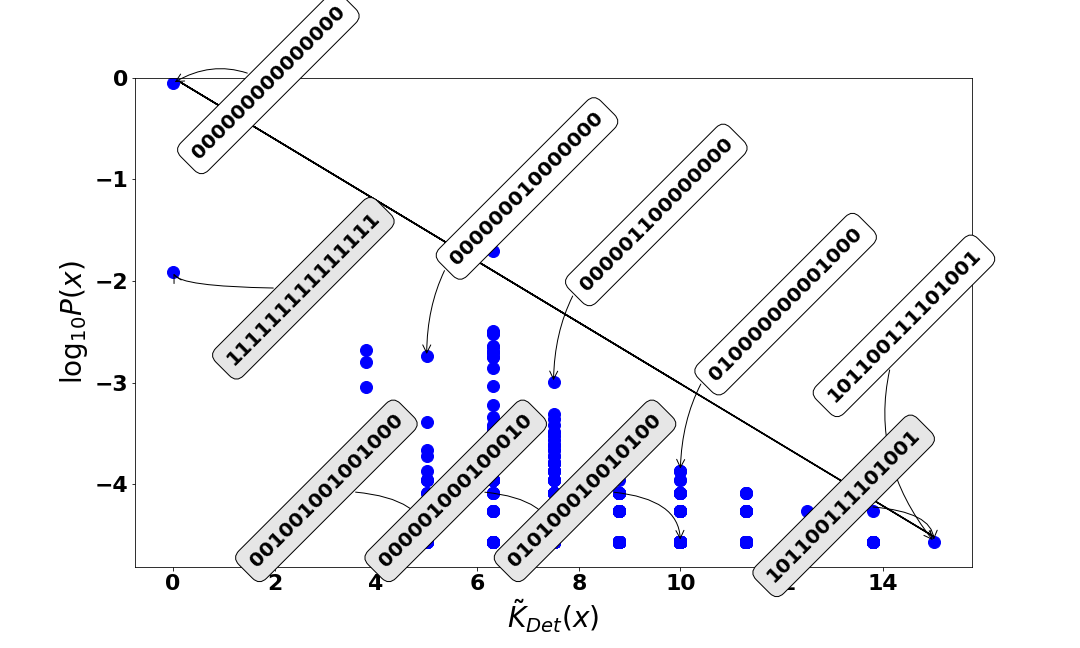}}
\end{center}
\caption{Simplicity bias (SB) in time series from a combined data set of different sectors from the World Bank Open Data, using different discretisation methods. Each blue dot is a different binary strings of length $n=15$ bits. (a) The High/Low (HL) method; (b) Up/Down (UD) method; and (c) Detrended up/down ($Det$) method. Panels (d), (e), and (f) show the same as (a), (b), and (c) but now with example binary strings displayed. SB is apparent in each case, with a decay in upper bound probability with increasing complexity. The upper bound (black line) prediction from Eq.\ (\ref{eq:SB}) describes the upper bound decay in the data quite well. }
\label{fig:scatter_combined}
\end{figure*}

\subsection{Theory}
The Kolmogorov complexity $K_U(x)$ of a string $x$ with respect to $U$,  is defined \cite{solomonoff1960preliminary,kolmogorov1965three,chaitin1975theory} as
\begin{equation}
K_U(x) = \min_{p}\{|p|: U(p)=x\}
\end{equation}
where $p$ is a binary program for a prefix optimal universal Turing machine (UTM) $U$ \cite{turing1936computable}, and $|p|$ indicates the length of the binary program $p$ in bits.  Due to the invariance theorem \cite{li2008introduction} for any two optimal UTMs $U$ and $V$, $K_U(x) = K_V(x)+O(1)$ so that the complexity of $x$ is independent of the machine, up to additive constants. Hence  we conventionally drop the subscript $U$ in $K_U(x)$, and speak of `the' Kolmogorov complexity $K(x)$. Informally, $K(x)$ can be defined as the length of a shortest program that produces $x$, or simply as the size in bits of the compressed version of $x$. If $x$ contains repeating patterns like $x=1010101010101010$ then it is easy to compress, and hence $K(x)$ will be small. On the other hand, a randomly generated bit string of length $n$ is highly unlikely to contain any significant patterns, and hence can only be described via specifying each bit separately without any compression, so that $K(x)\approx n$ bits. Other more expressive names for $K(x)$ are \emph{descriptional complexity}, \emph{algorithmic complexity}, and \emph{program-size complexity}, each of which highlight the idea that $K(x)$ is measuring the amount of information to describe or generate $x$ precisely and unambiguously. Note that Shannon information and Kolmogorov complexity closely are related \cite{grunwald2004shannon}, but differ fundamentally in that Shannon information quantifies the information or complexity of a random source, while Kolmogorov complexity quantifies the information of individual sequences or objects. More details on AIT can be found in refs.\ \cite{li2008introduction,calude2002information,gacs1988lecture}. 

Levin's coding theorem \cite{levin1974laws} states that
\begin{equation}
P(x) = 2^{-K(x)+O(1)}\label{eq:CD}
\end{equation}
where $P(x)$ is the probability that UTM $U$ generates output string $x$ on being fed random bits as a program (again we have dropped the subscript $U$). Thus, high complexity outputs have exponentially low probability, and simple outputs must have high probability. This is a profound result which links notions of data compression and probability in a direct way. $P(x)$ is also known as the \emph{algorithmic probability} of $x$.

As discussed in the Introduction, applying results from AIT is not straightforward, and typically results and complexity estimates must be approximated in various ways.  Coding theorem-like behaviour in real-world input-output maps was studied recently, leading to the observation of a  phenomenon called \emph{simplicity bias} (SB) \cite{dingle2018input} (see also ref.\ \cite{buchanan2018natural} for commentary on that work). SB is captured mathematically as
\begin{equation}
P(x)\leq 2^{-a\tilde{K}(x)-b}\label{eq:SB}
\end{equation}
where $P(x)$ is the (computable) probability of observing output $x$ on random choice of inputs, and $\tilde{K}(x)$ is the estimated Kolmogorov complexity of the output $x$: complex outputs from input-output maps have lower probabilities, and high probability outputs are simpler. The constant $a>0$ can often be estimated without recourse to sampling, but just by knowing or estimating the total number of different possible outputs \cite{dingle2018input}. Using $b=0$ is the default guess for this constant, but it can also be fit to the data via partial sampling. In this work we will use $b=0$ throughout.
The ways in which SB differs from Levin's coding theorem include that it does not assume UTMs, uses approximations of complexities, and for many outputs $P(x)\ll 2^{-K(x)}$. Hence the abundance of low complexity, low probability outputs \cite{dingle2020generic,alaskandarani2022low} is a signature of SB. 

A full understanding of exactly which systems will, and will not, show SB is still lacking, but the phenomenon is expected to appear in a wide class of input-output maps, under fairly general conditions. A main condition is that the map should be `simple' (technically of $O(1)$ complexity) to prevent the map itself from dominating over inputs in defining output patterns \cite{dingle2018input}. If an arbitrarily complex map was permitted, outputs could have arbitrary complexities and probabilities, and thereby remove any connection between probability and complexity.  If a pattern or output is generated via some `simple' computation of inputs to outputs, and there is bias (i.e.\ a strongly non-uniform probability distribution) then we expect to see SB. If the output patterns are merely uniform random strings of independent bits, then SB will not be observed. In the real world, output patterns may result from a mixture of these two (ie `computed' and randomised), and therefore may partially show SB. 

Many AIT applications \cite{li2008introduction} rely on approximations of Kolmogorov complexity via standard lossless compression algorithms, e.g.\ Lempel-Ziv \cite{lempel1976complexity,ziv1977universal}. In this work we follow this same practice, and in particular follow the slightly adjusted complexity measure $C_{LZ}(x)$ of ref.\ \cite{dingle2018input}  ie
\begin{equation}
C_{LZ}(x) =\begin{cases}
     \log_2(n), &  \hspace*{-0.3cm}  \text{$x=0^n$ or $1^n$}\\
    \log_2(n) [N_w(x_1...x_n) + N_{w}(x_n...x_1)]/2, & \hspace*{-0.2cm} \text{otherwise}
  \end{cases}\label{eq:CLZ}
\end{equation}
where $N_w(x)$ forms the basis for this complexity measure Lempel and Ziv \cite{lempel1976complexity}, and 
where the simplest strings $0^n$ and $1^n$ are separated out because  $ N_{w}(x)$ assigns complexity of 1 to the string 0 or 1, but complexity 2 to $0^n$ or $1^n$ for $n\geq2$, whereas the true Kolmogorov complexity of such a trivial string actually scales as $\log_2(n)$ for typical $n$, because one only needs to encode $n$. 
Having said that, the minimum possible value is $K(x)$$\approx$$0$ for a simple set, and so e.g.\ for binary strings of length $n$ we can expect $0 \lesssim K(x) \lesssim n$ bits. Because for a random string of length $n$ the value $C_{LZ}(x)$ is often much larger than $n$, especially for short strings, we scale the complexity so that $a$ in Eq.\ (\ref{eq:SB}) is set to $a=1$ via
\begin{equation}
\tilde{K}(x) = \log_2(M) \cdot \frac{ C_{LZ}(x) - \min_x (C_{LZ})}{\max_x (C_{LZ}) - \min_x (C_{LZ}) } \label{eq:Kscaled}
\end{equation}
where $M$ is the maximum possible number of output patterns in the system, and the min and max complexities are over all strings $x$ which the map can generate.  $\tilde{K}(x)$ is the approximation to Kolmogorov complexity that we use throughout. 
This scaling results in $0\leq \tilde{K}(x) \leq n$ which is the desirable range of values (within logarithmic terms), and also it results in $a=1$. If $M$ is not known, then it might estimated, but if it cannot be estimated then the complexity scaling will not be possible.

\begin{figure*}[htp]
\begin{center}
\subfigure[]{\label{fig:edge-a}\includegraphics[height=6.0cm,width=6.0cm]{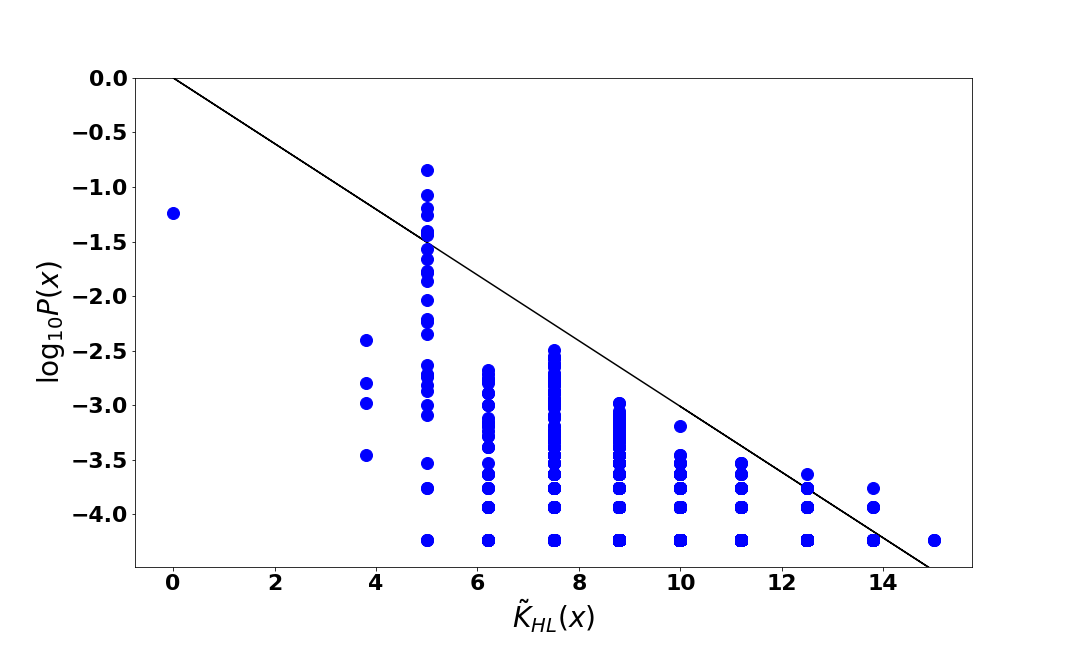}}
\subfigure[]{\label{fig:edge-a}\includegraphics[height=6.0cm,width=6.0cm]{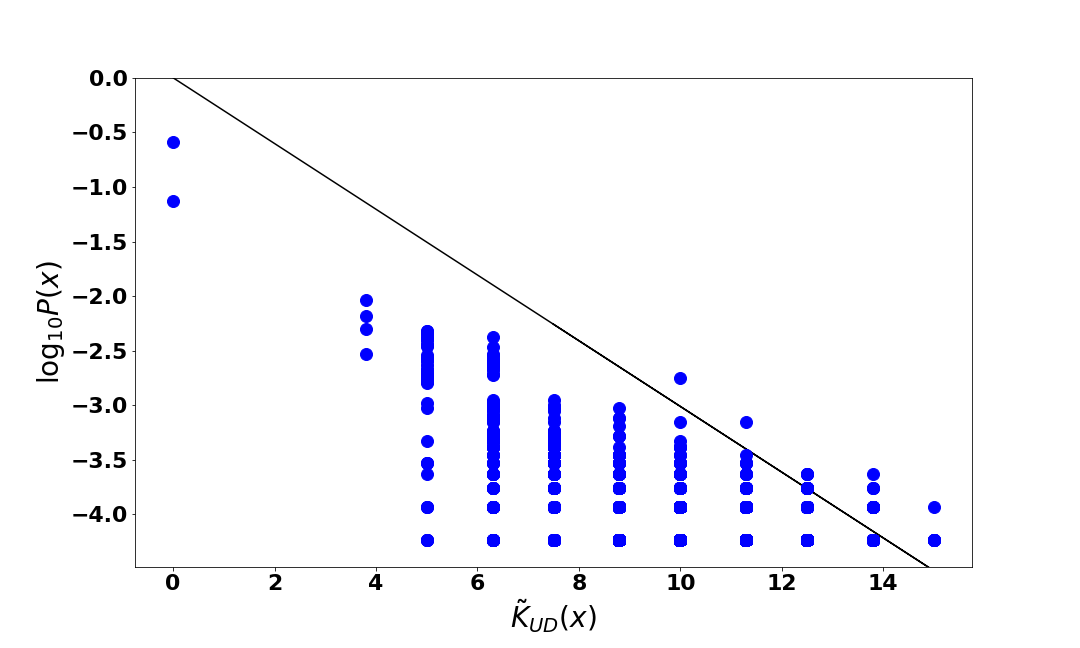}}
\subfigure[]{\label{fig:edge-a}\includegraphics[height=6.0cm,width=6.0cm]{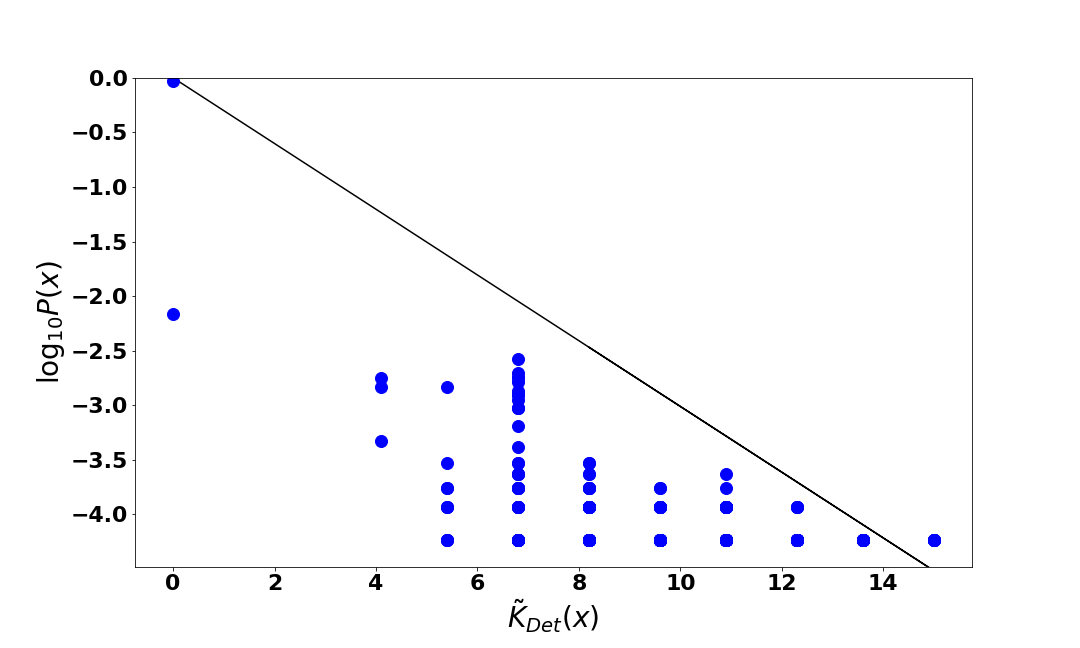}}
\subfigure[]{\label{fig:edge-a}\includegraphics[height=6.0cm,width=6.0cm]{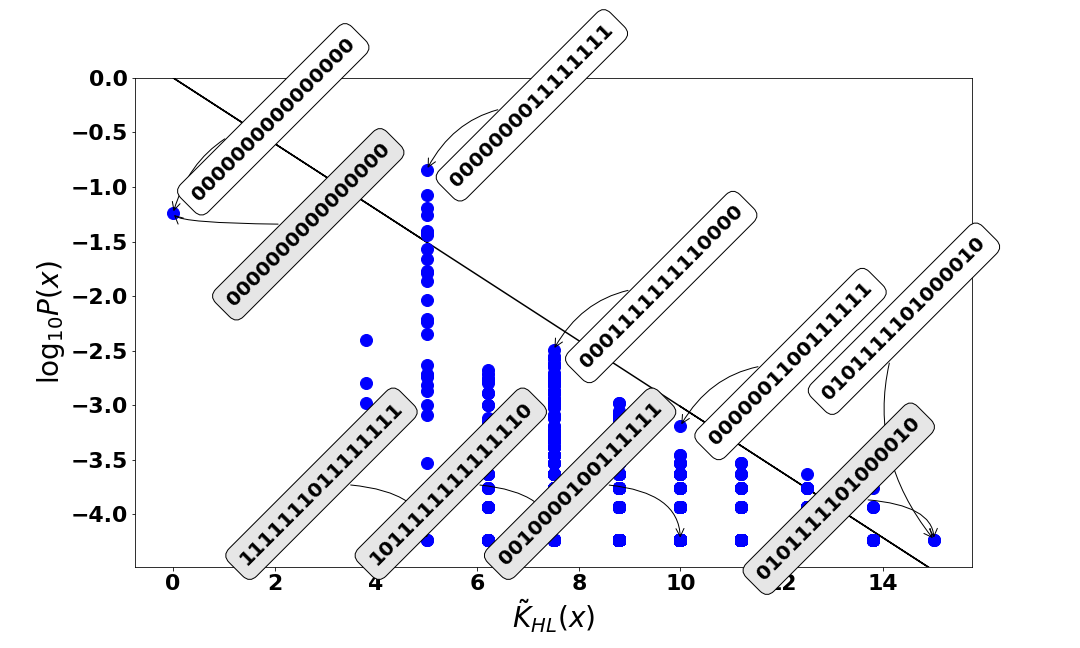}}
\subfigure[]{\label{fig:edge-a}\includegraphics[height=6.0cm,width=6.0cm]{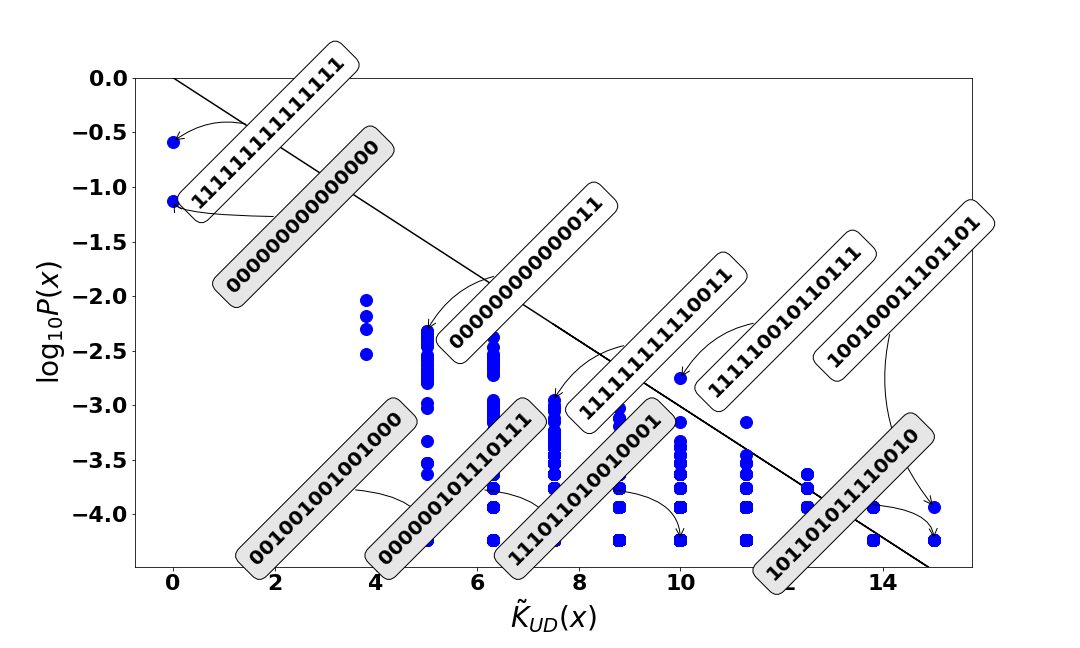}}
\subfigure[]{\label{fig:edge-a}\includegraphics[height=6.0cm,width=6.0cm]{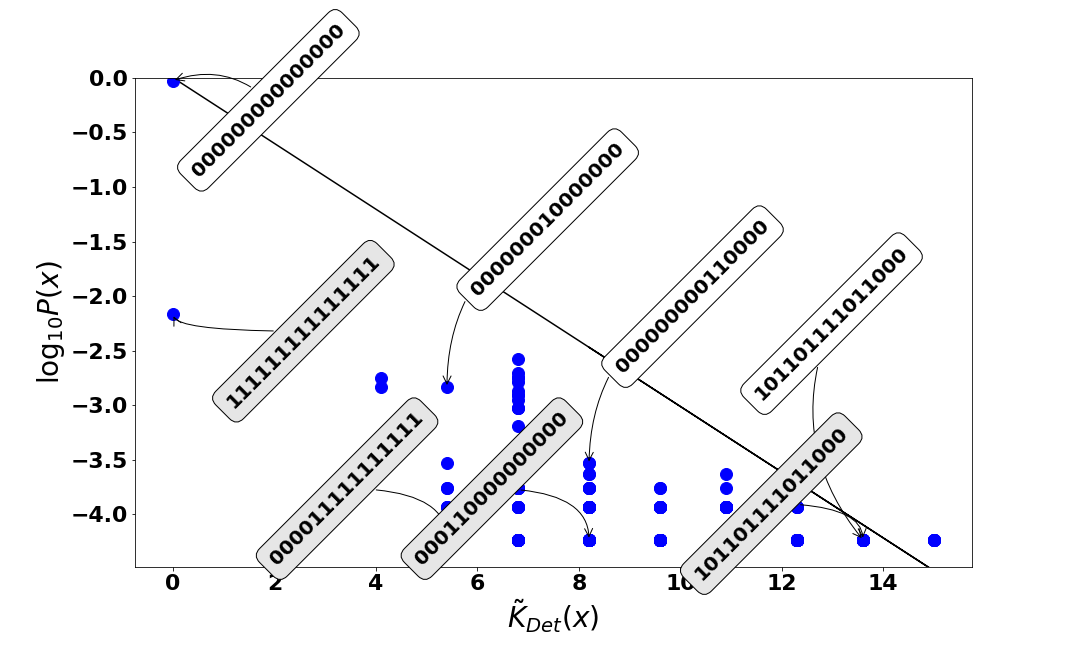}}
\end{center}
\caption{Simplicity bias (SB) in time series world health data, using different discretisation methods. Each blue dot is a different binary strings of length $n=15$ bits. (a) Low/High (LH) method; (b) Up/Down (UD) method; and (c) Detrended up/down (Det) method. (d)---(f) show the same as (a)---(c) but now with example binary strings displayed. SB is apparent in each case, with a decay in upper bound probability with increasing complexity. The upper bound (black line) prediction from Eq.\ (\ref{eq:SB}) describes the upper bound decay int he data quite well.}
\label{fig:scatter_worldhealth}
\end{figure*}

\subsection{Time series data}

To study simplicity bias (SB) in time series, we obtained series data from the World Bank Open Data website (\texttt{data.worldbank.org}, accessed Nov-Dec 2021), which gathers data of political, economic, and geographical interests, such as health, education, and energy statistics  for different countries. The choice of data set is somewhat important when studying SB, and we expect SB to be more pronounced for `natural' data sets. Expanding on this,  if a data set was chosen to contain particularly complex irregular series, then it is unlikely that SB would be observed, due to the intentional bias towards more complex series. The World Bank data was chosen because the time series contained therein were not chosen for any particular features of the series, but are a natural generic collection of data. 

The World Bank series data we obtained span a 50 year period 1971, 1972, 1973, ...  2020, and come from many countries. The sectors include health, development indicators, nitrous oxide emissions, employment records, economic monitors, energy and mining, and education statistics. From these data, we extracted series of length $n$ years, and in particular we chose the most resent run of $n$ consecutive series values which do not have  missing values. (The choice of the length of the series $n$ is discussed below.) Because many of these series contained missing data, the most recent series of $n$ values often did not end in 2020.  For example, to take a series of length $n=15$ years, if the period 2006, 2007, ..., 2020 had no missing values, then that period was used because it is the most recent. On the other hand if, say, data for the year 2012 was missing, then we took the next most recent consecutive run of data, i.e.\ a series from 1997 to 2011, etc. If a series had so many missing values that no sufficiently long series could be extracted, then the series was excluded from the study.

In this work we study mainly two collections of data. Firstly, a combined data set amalgamating series from all the sectors mentioned above. The reasons for combining the data are primarily to obtain a larger data set to study, but also to reduce the likelihood of observing an idiosyncratic set of series patterns particular to  one sector.  Secondly, we focus on just the world health data. The reason for this choice was that this set of series was well populated with data, and also to test if indeed any idiosyncratic patterns preferences appeared which might disrupt SB. Note than many of the sectors data did not have many series with sufficient quality.

\subsection{Discretising time series}

SB and results from AIT are most easily applied in the context of binary string patterns, so the real-valued World Bank time series were each converted into binary strings. Here we study these resulting binary strings and their probabilities.  Denote by $s(t)$ the real-valued time series, where $t$ represents the year. We then discretise $s(t)$ to a binary string $x(t)\in\{0,1\}^n$ of length $n$. In principle, there are many different ways this discretisation can be done:  
One option is to first calculate the mean value of the series $s(t)$ over some given time period, and then replace $s(t)$ with 1 if $s(t)>$mean($s$), and replace $s(t)$ with 0 if $s(t)\leq$mean(s). As an example, if the $n=4$ series values were 23.5, 28.1, 19.5, 18.3, then the mean of these is 22.35 and hence the binary string would be $x=1100$ because the first two values are above the mean, and the second two below. This is a kind of high/low (HL) discretisation. We will denote the complexity of the binary string $x$ resulting from this  HL method as $\tilde{K}_{HL}(x)$, which comes from applying the complexity measure of Eq.\ (\ref{eq:Kscaled}) to the HL binary string. A second option to use the `up/down' method \cite{hansen2001model,willbrand2005identifying,dingle2018input}, where if a series increases $ds/dt> 0$ from year $t$ to $t+1$ then $s(t)$ is replaced by a 1, and $s(t)$ is replaced by a 0 if $dx/dt \leq 0$. We will denote the complexity of the binary string resulting from this up/down (UD) method as $\tilde{K}_{UD}(x)$. In this work, we additionally make a slight adjustment to this UD method, to make a third option: first performing a linear detrending of the series, i.e.\ subtracting a linear fit from the series $s(t)-(\alpha t + \beta)$ for fitted constants $\alpha$ and $\beta$, and then converting the resulting up-down series into a binary string according to the UD method. We call this detrended up/down method $Det$.  The reason for introducing this third option is that it may be that many time series have a monotonic increasing or decreasing trends, leading to many UD discretised series becoming the rather trivial patterns 111...111 or 000...000. We will denote the complexity of the binary string resulting from this detrended up/down  method as $\tilde{K}_{Det}(x)$. SB studies typically use only one disretisation procedure, but here we use three here and compare them.

To illustrate the discretisation process and resulting binary strings, we use HIV infection series from the world health data. Figure (\ref{fig:blueyellow})(a) shows a series discretised by the HL method where yellow sections denote high (above mean) $s(t)$ values which are converted to 1, and blue segments for low $s(t)$ values, which are converted to 0. The resulting binary string from the series in the figure panel is $x=$ 1100000000001111111110111110000 which has length $n=31$ bits. Note that the segments $s(t)$ to $s(t+1)$ are coloured according to the value of $s(t)$, so it may be that, for example, some yellow segments appear to dip into the lower half, but this is merely due to the fact that $s(t)$ is high while  $s(t+1)$ is low. Figure (\ref{fig:blueyellow})(b) illustrates the UD method, where the sign of the slope $ds/dt$ is used to define the binary string. The resulting binary string is $x=$ 11110001100110011101 with $n$=20.
Figure (\ref{fig:blueyellow})(c) shows the analogous plot for the $Det$ method, which is fundamentally quite similar to the UD method. The resulting binary string is $x=$ 00000010110110101010 with $n=20$.

Note that, as with any discretisation process, many different real-valued series will correspond to a given binary pattern. Hence we are not attempting to predict a particular real-valued  series, but rather predicting the probability of a given discretised series pattern. 

\subsection{Pattern probability bounds from complexities}

Empirically, SB has been observed in systems with small output patterns, e.g.\ of $n\approx50$ bits \cite{dingle2018input} and even with $\approx$10 bits \cite{delahaye2012numerical,soler2014calculating} in some cases. However, both in theory and in practice SB is expected to be more clearly observed for systems where the size of the output pattern/string $n$ is large. Because there are up to $2^n$ different binary series of length $n$ bits, obtaining decent frequency statistics for computing the probability $P(x)$ of $2^n$ patterns requires a large amount of data if $n$ is large. Hence there is a trade-off between choosing a larger or smaller $n$, depending on the amount of data available. Given our data sets   have around $10^3$ to $10^4$ series, we chose $n=15$ bits for our current analysis (but see the Appendix for other sizes $n$).

Two main predictions are to be tested with the data: firstly, that an inverse relation of complexity and probability is observed, i.e.\ SB; and secondly that the upper bound on the data follows that of Eq.\ (\ref{eq:SB}). These two aspects are distinct in that it is possible to observe the first without the second, i.e.\ the slope  $a$ and intercept $b$ may not be accurately predicted, even though the SB phenomenon is observed. If the data do follow the bound of Eq.\ (\ref{eq:SB}), then this would constitute a form of series pattern forecasting without fitting to data, i.e.\ \emph{a priori} forecasting. 

We begin with probability predictions for a combined data set amalgamated from various areas namely health, education, energy, and economics, as described above. This combined data set contains $\sim$36,000 series. Figure \ref{fig:scatter_combined}(a) results from using the HL method. SB is apparent in this plot, and the upper bound prediction (black line) describes the upper bound quite closely. This prediction bound is based on no fitting to the data, except to know the mean series value which is used to determine whether a value is high (H) or low (L). In a scenario where a mean value can be known or estimated \emph{a priori} then even this value need not be fit. For example, the mean might be estimated \emph{a priori} from physical constraints.

Turning to the UD method,  in Figure \ref{fig:scatter_combined}(b) SB is observed but the pattern is somewhat noisier, especially around $\tilde{K}_{UD}(x)\approx$13 where there is a cluster of points above the prediction line. These points of overabundance could be due to some sector-specific pattern preference, or could be due to added randomness in the series (e.g.\ from measurement errors) which would tend to spoil the SB. Figure \ref{fig:scatter_combined}(c) shows a plot of the same data but for the $Det$ method, in which SB is quite clear, with a clear relation decay in probability with increasing complexity. The upper bound describes the data quite closely, but not as well as for the HL method. 

Figures \ref{fig:scatter_combined}(d)---(f) show the same plots as Figure \ref{fig:scatter_combined}(a)---(c) but now with some binary strings displayed for illustrative purposes. Specifically, examples of the highest and lowest probability strings for different complexity values as plotted. It appears that one difference between high and low probability strings of the same complexity is that lower probability strings have more 0/1 or 1/0 changes in them. We studied this observation in a separate study \cite{alaskandarani2022low}. The observation that strings with the same complexity can have very different probabilities was made earlier \cite{dingle2018input}, where it was reported that \emph{most} output binary string patterns fall far below the upper bound black line, i.e.\ $P(x)\ll 2^{-a\tilde{K}(x)-b}$. Counterintuitively, at the same time most of the probability mass is accounted for by points which are close to the bound \cite{dingle2018input}. Thus, for a randomly selected series, the corresponding binary string pattern is likely to be close to the bound with high probability \cite{dingle2020generic}, i.e. $P(x)\approx 2^{-a\tilde{K}(x)-b}$. In ref.\ \cite{dingle2020generic} this phenomenon of low-complexity, low-probability outputs was studied  analytically and numerically, and it was suggested that such patterns are those that are intrinsically not very complex, but are difficult to make for particular system. Nonetheless, this phenomenon is still not fully understood, and in a sense these points far below the bound represent a failing of our probability predictions. Hence a better grasp of why such patterns arise and which patterns have this property would help in improving probability predictions (see more in ref.\ \cite{alaskandarani2022low}). 
 
In Figure \ref{fig:scatter_worldhealth} we study SB in one sector only --- world health --- to see if SB persists for particular sectors (as opposed to being a property of aggregate series only). 
The general trends of these graphs are similar to the combined data set of Figure \ref{fig:scatter_combined}, with  SB observed in all three methods. In particular, the UD method shows more pronounced SB as compared to in the combined data, and the bound is followed quite closely, but the slope is not quite as accurate. Additionally, the upper bound prediction agrees with the data bound quite well. Predictions using the $Det$ method are not as compelling as the HL and UD methods. In the Appendix we show complexity-probability graphs for another sector,  world development indicators, and also compare the graphs when using $n=5$, $n=10$, $n=15$ and $n=20$.

\section{Complexity histograms}

The distribution of complexity values $P(\tilde{K}(x)=k)$ for a data set depends on (a)  the probability of each series pattern with complexity $k$ bits, and (b) the number of patterns with complexity $k$ bits. Hence the SB bound of Eq.\ (\ref{eq:SB}) does not necessarily imply that the average complexity of randomly selected series will be low, because there may be lots of complex strings whose combined probability mass is large even if individually they have low probability. According to SB, there is an exponential decay in probability with increasing complexity $k$, but on the other hand it is known that the number of patterns with complexity $k$ increases exponentially with $k$ \cite{li2008introduction}. If these two exponential trends are balanced, then the resulting distribution of complexities may be roughly uniform. However, if there lower numbers  of strings with complexity $k$, for example, then the distribution of complexities may tip either to higher or lower values.

\begin{figure*}[htp]
\begin{center}
\subfigure[]{\label{fig:edge-a}\includegraphics[height=6.0cm,width=6.0cm]{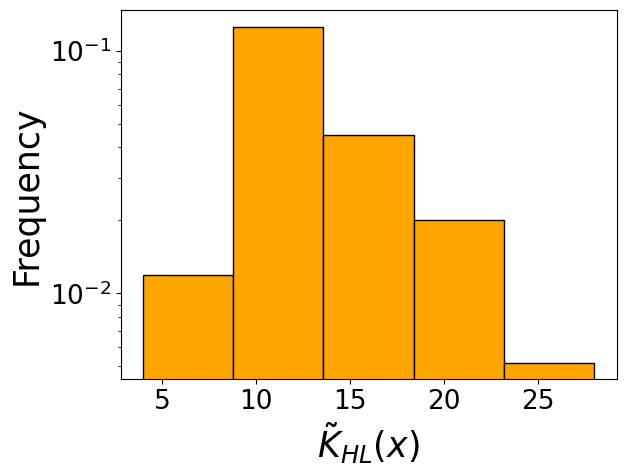}}
\subfigure[]{\label{fig:edge-a}\includegraphics[height=6.0cm,width=6.0cm]{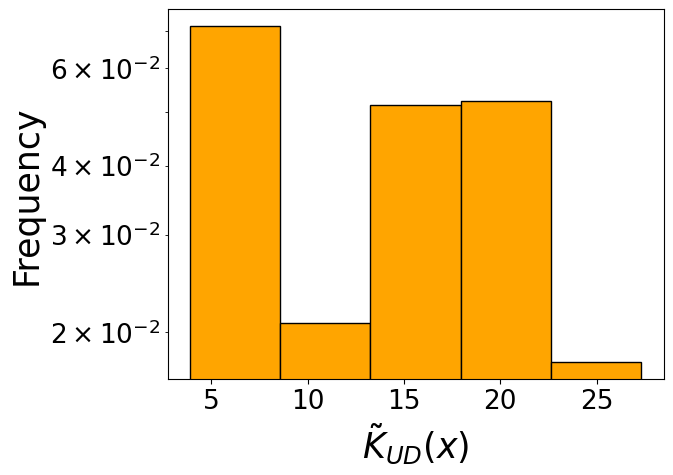}}
\subfigure[]{\label{fig:edge-a}\includegraphics[height=6.0cm,width=6.0cm]{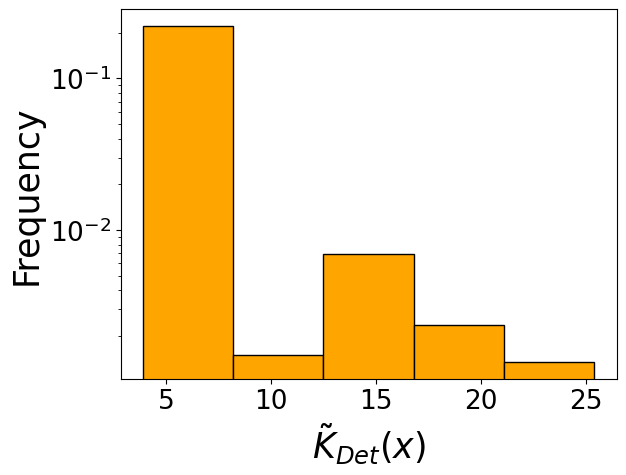}}
\end{center}
\caption{The distribution complexity value for the world health data, for (a) HL, (b) UD, and (c) $Det$ methods.  In all three methods there is a bias towards low complexity values, with mean complexity values $\mu=$13.7, 13.2, and 4.5 respectively, much lower than $\mu=$21.1 bits for completely random bit strings. }
\label{fig:distn_K}
\end{figure*}

In Figure \ref{fig:distn_K} we show the distribution of complexity values for each method, using the the world health data with $n=15$ as above. For all three methods, there is a noticeable bias towards lower complexities. Numerically,  the mean complexity for the HL histogram is $\mu=$13.7 with standard deviation $\sigma$=4.2, for UD $\mu=$13.2 with $\sigma=$7.3, for $Det$ $\mu=$4.5 with $\sigma=$2.7. For comparison, we  the mean complexity of all possible binary strings of length 15 bits (equivalent to a uniform distribution over strings) is $\mu=$21.1 with standard deviation $\sigma=$2.34. All three mean values are much less than the mean for purely random strings. So, in this case of time series, not only do individual series show SB, but the distribution is also strongly biased towards simpler patterns.

\section{Predicting which of $P(x)$ or $P(y)$ is higher}

We now return to studying the probability $P(x)$ of observing a given series pattern $x$.
Given two binary series patterns $x$ and $y$, can we predict which has a higher probability without relying on empirical frequency count data? In many cases the primary interest is to guess whether $P(x)>P(y)$ or $P(x)<P(y)$, rather than trying to guess the exact values of $P(x)$ and $P(x)$. For example in investing, one might want to know which series pattern to bet on, $x$ or $y$. 
Notice that to use Eq.\ (\ref{eq:SB}) for this prediction,  both the constants $a$ and $b$ are irrelevant, because whether $2^{-a\tilde{K}(x)-b}$  or $2^{-a\tilde{K}(y)-b}$ is larger does not depend on these constants. The prediction just depends on the relative size of $\tilde{K}(x)$ and $\tilde{K}(y)$. So even if we could not estimate $a$ or $b$ accurately, we may still be able to determine which outcome $x$ or $y$ is more likely. 

For the World Bank data sets, we experimented with this prediction, while also accommodating the situation where $\tilde{K}(x)=\tilde{K}(y)$. In more detail, for a given data set, we chose two real valued series at random from the dataset, then converted them both to binary strings $x$ and $y$ according to the HL method.  We then predict
\begin{eqnarray}
P(y)>P(x) &\texttt{ if }& \tilde{K}(y)<\tilde{K}(x)\\
P(y)<P(x)& \texttt{ if } &\tilde{K}(y)>\tilde{K}(x)\\
P(y)>P(x) &\texttt{ w/prob. 0.5 if } &  \tilde{K}(y)=\tilde{K}(x)
\end{eqnarray}
For the last condition, a random number (e.g.\ coin flip) is drawn to guess whether $P(y)>P(x)$ or $P(y)<P(x)$ each with 50\% probability. It may seem more natural to predict $P(y)=P(x)$ when $\tilde{K}(y)=\tilde{K}(x)$, but following this suggestion will lead to erroneous predictions most of the time. The reason is that there are very many unique probabilities values so that it is highly unlikely that $P(y)=P(x)$ exactly, whereas the complexities are far more coarsely measured, and hence coincidences of complexity values are much more likely. According to this prediction protocol, if complexity had no role in modulating probabilities, then in theory the null success rate of correctly guessing which of $P(x)$ or $P(y)$ is higher should be 50\%. Testing  this prediction protocol with the time series data sets studied above and choosing 10,000 random  ($x$,$y$) pairs yielded the following success results:\\

\noindent
Combined Data, HL method, success rate 81\%\\
Combined Data, UD method, success rate 84\%\\
Combined Data, Det method, success rate 59\%\\

\noindent
World Health Data, HL method, success rate 75\%\\
World Health Data, UD method, success rate 88\%\\
World Health Data, Det method, success rate 55\%\\

So we see that without fitting to the data, and simply using the relative size of the complexity estimates as a predictor, success rates well above random null model can be achieved (all $p-$values$<10^{-3}$). Of the three methods, the UD method had the highest success rate followed quite closely by the HL method. The $Det$ performed quite badly, only marginally above the baseline 50\% prediction accuracy. This failure can be rationalised by noting that in Figure \ref{fig:scatter_combined}(c) and Figure \ref{fig:scatter_worldhealth}(c) the vast majority of the probability mass is taken up by only 2 or 3 data points, for which there is little local relation to complexity. 
Nonetheless, the fact that we can predict which series is more likely with the HL and UD methods with high  accuracy, while utilising only minimal details of the map, is quite surprising and potentially useful in applied probability settings. 

\section{Comparison to Brownian motion}

Brownian motion is commonly used as a baseline null model of time series trajectory. Given that we have observed SB in real time series, it is interesting to study SB in artificially generated Brownian motion also. In ref.\ \cite{dingle2018input} (Supplementary Information) SB was briefly studied  both numerically and analytically using the HL discretisation of random walks (the discrete analogue of Brownian motion), and evidence of SB was found.

\begin{figure*}[htp]
\begin{center}
\subfigure[]{\label{fig:edge-a}\includegraphics[height=6.0cm,width=6.0cm]{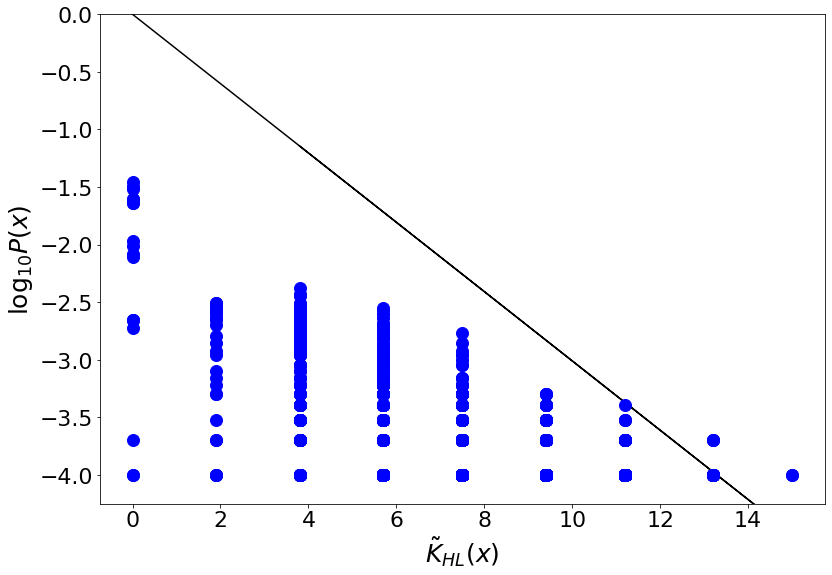}}
\subfigure[]{\label{fig:edge-a}\includegraphics[height=6.0cm,width=6.0cm]{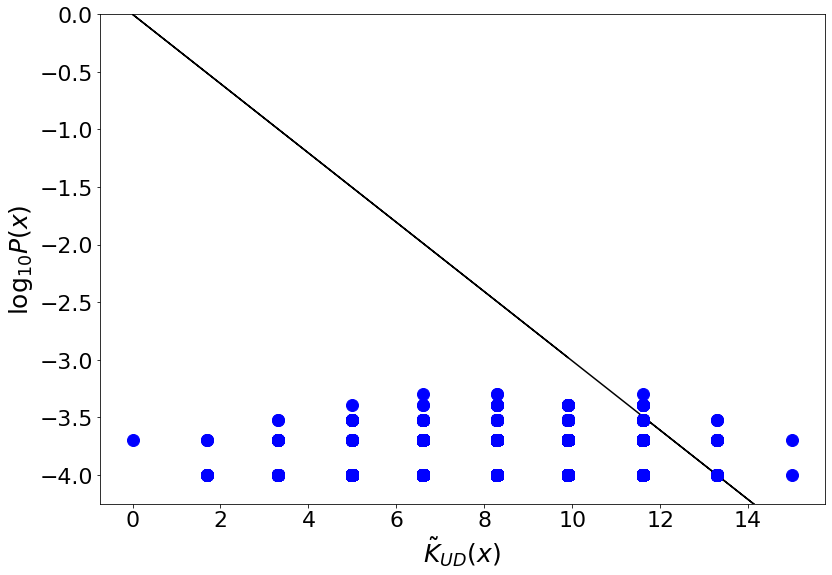}}
\subfigure[]{\label{fig:edge-a}\includegraphics[height=6.0cm,width=6.0cm]{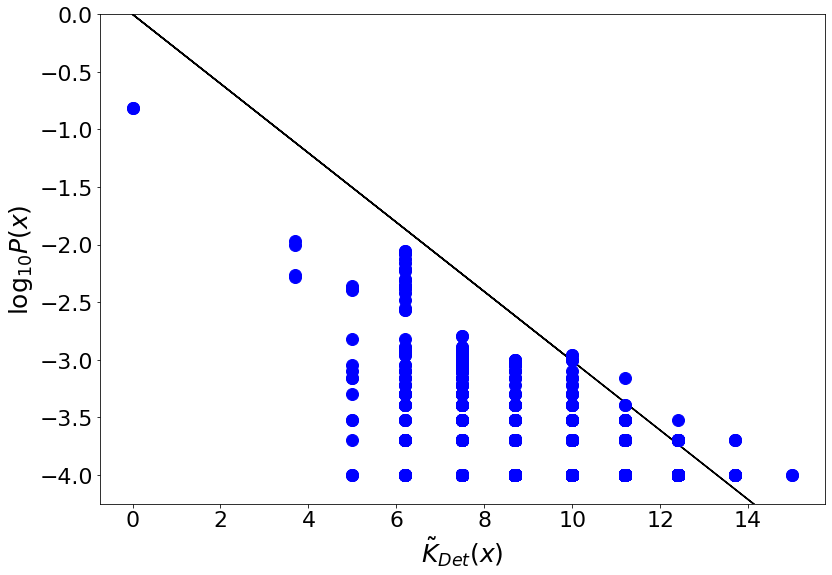}}
\end{center}
\caption{Examining simplicity bias (SB) in artificial Brownian motion series. (a) The HL method shows some SB, but it is not as clear as for the real-world data; (b) the UD method shows no SB at all; (c) the $Det$ method shows clear SB.}
\label{fig:Bm}
\end{figure*}

We numerically generated 10,000 Brownian motion series with $n=15$, and Figure \ref{fig:Bm} shows the resulting probability-complexity plots. Starting with the HL method in panel (a), there is certainly some SB observed, but the bound is not followed closely, and also the slope of the decay in probability is less steep. Turning to panel (b), the SB has disappeared completely, hence this marks a strong contrast between the Brownian motion and real-world data. The absence of SB is expected in fact, because in Brownian motion the increases and decreases at each time step is determined by Gaussian noise and independent of past values, and hence the UD profile is defined essentially by a uniformly random string of 0s and 1s. Therefore there is no bias, and no SB. Finally in (c), the $Det$ method shows quite clear SB, somewhat in contrast to the real-world combined and health data sets for which $Det$ showed less clear SB. We conclude that SB is present to some extend in artificially generated Brownian motion series, but it is less clear and consistent than in the real time series from the World Bank data.

Another baseline model for time series is white noise, i.e.\ uncorrelated Gaussian random variables. With a time series of white noise, the HL method will not yield SB because the binary string resulting from will just be a random strings of 0s and 1s, with no bias and no SB. With the UD method, again it is easy to see that no bias and no SB would be observed. With the $Det$ it is possible that some SB would be observed because the detrending may introduce some  structure to the series. We will not investigate this trivial white noise series further.

\section{Discussion}

We have numerically studied simplicity bias (SB) and \emph{a priori} forecasting in discretised real-world time series data sets. We studied different discretisation procedures also. Our main finding in this short investigation is that we report observing SB in these series, and that it appears we can make non-trivial predictions regarding the probability of different time series patterns, using very little or no details of the system, or fitting to historical data. In this manner, we present a kind of forecasting without historical data, in which the different patterns or shapes of the series can be predicted (but not the actual numerical series values themselves). Of course we are not claiming that accurate quantitative predictions can be made for series without fitting to trends in historical data, nor even that very accurate predictions of discretised patterns can be made. But we are claiming that even without recourse to such training data, non-trivial forecasts about patterns of trajectories can be made. We also showed that predictions regarding which of two randomly chosen series has higher probability can be made with a high success rate of $\sim$80\%. With our data sets, these success rates were high for the HL and UD methods, but not the $Det$ method. Based on this small study, it appears that the HL and UD methods are the most suitable for observing SB, in the sense that the real world data sets for these two methods (and the prediction for which is string more likely) more closely followed the expectations of SB.

Why does SB appear in these series? The original AIT coding theorem was proved for Turing machines, so what is the connection to, for example, health and education statistics which arise from economic and social factors, among others? If we view time series as the result of some kind of computation, potentially involving many separate factors but nonetheless together acting to `compute' a series pattern then the connection is more clear. In economics, for example, it is widely believed that the levels of unemployment and other financial metrics are the result of various other factors which via mathematical equations and `computations' determine (at least roughly) economic outcomes. For the SB predictions of ref.\ \cite{dingle2018input} to apply, it is not required that a Turing machine produce the patterns. Instead, the premise of ref.\ \cite{dingle2018input} is that behaviour of real-world computational processes can be \emph{bounded} by the behaviour of a Turing machine. These bounds can be used for prediction. From another angle, because the SB relation is so strong, i.e.\ an exponential decay, then even if it the theory only applies weakly to a system, we can still expect to see some evidence of SB.

Although technically uncomputable, Kolmogorov complexity is essentially merely a measure of the size in bits of the compressed version of a data object such as a binary pattern. Hence, Vitanyi \cite{vitanyi2020incomputable,vitanyi2013similarity} points out that because naturally generated data is unlikely to contain pseudo random complexities like $\pi$, the true complexity is unlikely to be much shorter than that achievable by every-day compressors. 
Further, it is worth noting that the SB bound is specifically relevant in the computable (i.e.\ non-UTM) setting. Indeed, Zenil et al \cite{zenil2019coding} have numerically studied Levin's coding theorem behaviour and found it persists for different types of computing systems with different levels of computational power, again indicating that UTMs are not strictly necessary. Therefore we can expect a general probability-complexity relation to hold outside the abstract setting in which it was initially proved.

Further reasons to understand why results based on  the AIT coding theorem should work in real-world applications can be found in information theory, research developed largely independently of Levin's work. The fundamental connection between probability and data compression has also been studied by Cover \cite{cover1974universal}, Langdon \cite{langdon1983note}, and Rissansen \cite{rissanen1984universal}. Since then, different communities --- e.g.\ information theory \cite{feder1992universal,cover2006elements}, optimal gambling strategies \cite{feder1991gambling}, and password guessing \cite{merhav2019universal} --- have studied and exploited the probability-compression connection without utilising Kolmogorov complexity \emph{per se}, but instead Lempel-Ziv style compression approaches. In a review, Merhav and Feder \cite{merhav1998universal} surveyed results in the area known as \emph{universal prediction} and explicitly point to $2^{-LZ(x)}$ as an effective universal probability assignment for prediction based on the results of ref.\ \cite{plotnik1992upper} and others, where $LZ(x)$ is the Lempel-Ziv compression complexity measure, essentially the same as we use here.  These results all support the use of these abstract information theoretic arguments in practical predictions contexts.

A weakness of our predictions is that they only constitute an upper bound on the probabilities, many output trajectory patterns $x$ fall far below their respective upper bounds. Following the hypothesis from \cite{dingle2020generic}, these low-complexity low-probability outputs are presumably patterns which the function or map finds `hard' to make, yet are not intrinsically very complex. Further, the presence of these low-complexity low-probability patterns may indicate the non-universal power of the map \cite{dingle2020generic,alaskandarani2022low}. We suggest that this phenomenon should be probed thoroughly.

Our work here is closely related to that of Ryabko et al \cite{ryabko2016compression,ryabko2009compression}, where real-world time series were forecasted using compression techniques. In ref.\ \cite{ryabko2016compression}  time series were extrapolated by choosing the future pattern which is most compressible, given the historical data pattern of some series. Our work differs from, and expands on, this area of research by not fitting to historical data patterns at all, and also by employing SB and algorithmic probability ideas. Also closely related is the work of Zenil and Delahaye \cite{zenil2011algorithmic}, in which financial time series were studied in terms of algorithmic probability. However, that study was somewhat different in that the aim was not forecasting. Fink et al \cite{fink20071} also studied time series in terms of algorithmic information content and data compression, including natural data from microarray gene expression. Again, while related to our study, the purpose of that study was nor forecasting \emph{per se}.

Our work is also related to the Minimum Description Length (MDL) principle \cite{rissanen1978modeling}, which is also fundamentally related to ideas from AIT. Indeed, MDL is partly an attempt to apply AIT ideas in practical situations, and so very much in the same vein as our work. In contrast however, MDL is most commonly used in model selection \cite{grunwald2000model,hansen2001model} as opposed to forecasting, as we do here.

Many future research directions remain open. These include: making explicit computations in the case of simple linear dynamical systems (linear autonomous/nonautonomous ODEs, linear discrete-time systems, linear stochastic differential equations), considering prototypical chaotic dynamical systems  and adding a small noise to the data and see how the results are affected. Also, exploring AIT using kernel methods in similar spirit of \cite{bach2022information} where one would consider SB and Kolmogorov complexity using kernel methods. An interesting direction would be to consider the question of kernel design from the point of view of AIT and considering the question of choosing kernels that lead to low complexity.
Another direction is to study probability predictions from 
different discretisation methods, e.g.\ combining HL and UD estimates. Finally, using a larger alphabet than merely binary 0/1 could be explored (cf.\ ref.\ \cite{ryabko2016compression}).
The methods we describe are general and can apply to other prediction systems. In future work it would be interesting to combine universal prediction methods like SB with other machine learning methods to predict time series (e.g. \cite{hamzi2021learning} where one would be interested in why this sort of method is efficient), potentially aiding or improving prediction (see also early \cite{schmidhuber1995discovering,schmidhuber1997shifting} and recent \cite{valle2018deep,zenil2019causal,hernandez2020algorithmic} studies on Kolmogorov complexity connections to machine learning). An important direction may be to use the SB bound as a prior in the Bayesian sense, as Hutter has described  \cite{hutter2007universal,hutter2004universal}, and even in a general theory of learning \cite{hutter2004universal}.

Finally, this work could also help introduce ways to define domains of applicability of some machine learning methods since we are able to make predictions without training and therefore methods based on SB could be viewed as a ``lower bound'' on ML methods.

\vspace{0.5cm}
\noindent
{\bf Acknowledgements:} KD acknowledges financial support from a Faculty Seed Grant case number 253571 awarded by the Gulf University for Science and Technology. BH acknowledges partial support by the Air Force Office of Scientific Research under MURI award number FA9550-20-1-0358 (Machine Learning and Physics-Based Modeling and Simulation).

\vspace{0.5cm}
\noindent
{\bf Data and code availability}: All the data for this study are available from the World Bank Open Data website \texttt{data.worldbank.org}. The code for this study is available at\\ \texttt{github.com/rafiq/Patterns-in-Real-World-Time-Series-Data}.

\noindent

\section{Bibliography}
\bibliographystyle{unsrt}

\begin{thebibliography}{}

\end{thebibliography}


\begin{thebibliography}{10}

\bibitem{solomonoff1960preliminary}
R.~J. Solomonoff.
\newblock A preliminary report on a general theory of inductive inference
  (revision of report v-131).
\newblock {\em Contract AF}, 49(639):376, 1960.

\bibitem{kolmogorov1965three}
A.N. Kolmogorov.
\newblock Three approaches to the quantitative definition of information.
\newblock {\em Problems of information transmission}, 1(1):1--7, 1965.

\bibitem{chaitin1975theory}
Gregory~J Chaitin.
\newblock A theory of program size formally identical to information theory.
\newblock {\em Journal of the ACM (JACM)}, 22(3):329--340, 1975.

\bibitem{li2008introduction}
M.~Li and P.M.B. Vitanyi.
\newblock {\em An introduction to Kolmogorov complexity and its applications}.
\newblock Springer-Verlag New York Inc, 2008.

\bibitem{levin1974laws}
L.A. Levin.
\newblock Laws of information conservation (nongrowth) and aspects of the
  foundation of probability theory.
\newblock {\em Problemy Peredachi Informatsii}, 10(3):30--35, 1974.

\bibitem{solomonoff2003kolmogorov}
Ray~J Solomonoff.
\newblock The kolmogorov lecture the universal distribution and machine
  learning.
\newblock {\em The Computer Journal}, 46(6):598--601, 2003.

\bibitem{sep-simplicity}
Alan Baker.
\newblock {Simplicity}.
\newblock In Edward~N. Zalta, editor, {\em The {Stanford} Encyclopedia of
  Philosophy}. Metaphysics Research Lab, Stanford University, {W}inter 2016
  edition, 2016.

\bibitem{hansen2001model}
Mark~H Hansen and Bin Yu.
\newblock Model selection and the principle of minimum description length.
\newblock {\em Journal of the American Statistical Association},
  96(454):746--774, 2001.

\bibitem{cilibrasi2005clustering}
R.~Cilibrasi and P.M.B. Vit{\'a}nyi.
\newblock Clustering by compression.
\newblock {\em Information Theory, IEEE Transactions on}, 51(4):1523--1545,
  2005.

\bibitem{ferragina2007compression}
P.~Ferragina, R.~Giancarlo, V.~Greco, G.~Manzini, and G.~Valiente.
\newblock Compression-based classification of biological sequences and
  structures via the universal similarity metric: experimental assessment.
\newblock {\em BMC bioinformatics}, 8(1):252, 2007.

\bibitem{avinery2019universal}
Ram Avinery, Micha Kornreich, and Roy Beck.
\newblock Universal and accessible entropy estimation using a compression
  algorithm.
\newblock {\em Physical review letters}, 123(17):178102, 2019.

\bibitem{vitanyi2013similarity}
Paul~MB Vit{\'a}nyi.
\newblock Similarity and denoising.
\newblock {\em Philosophical Transactions of the Royal Society A: Mathematical,
  Physical and Engineering Sciences}, 371(1984):20120091, 2013.

\bibitem{dingle2018input}
Kamaludin Dingle, Chico~Q Camargo, and Ard~A Louis.
\newblock Input--output maps are strongly biased towards simple outputs.
\newblock {\em Nature communications}, 9(1):761, 2018.

\bibitem{dingle2020generic}
Kamaludin Dingle, Guillermo~Valle P{\'e}rez, and Ard~A Louis.
\newblock Generic predictions of output probability based on complexities of
  inputs and outputs.
\newblock {\em Scientific reports}, 10(1):1--9, 2020.

\bibitem{johnston2022symmetry}
Iain~G Johnston, Kamaludin Dingle, Sam~F Greenbury, Chico~Q Camargo,
  Jonathan~PK Doye, Sebastian~E Ahnert, and Ard~A Louis.
\newblock Symmetry and simplicity spontaneously emerge from the algorithmic
  nature of evolution.
\newblock {\em Proceedings of the National Academy of Sciences},
  119(11):e2113883119, 2022.

\bibitem{dingle2022predicting}
Kamaludin Dingle, Javor~K Novev, Sebastian~E Ahnert, and Ard~A Louis.
\newblock Predicting phenotype transition probabilities via conditional
  algorithmic probability approximations.
\newblock {\em bioRxiv}, 2022.

\bibitem{delahaye2012numerical}
J.P. Delahaye and H.~Zenil.
\newblock Numerical evaluation of algorithmic complexity for short strings: A
  glance into the innermost structure of algorithmic randomness.
\newblock {\em Appl. Math. Comput.}, 219:63--77, 2012.

\bibitem{soler2014calculating}
Fernando Soler-Toscano, Hector Zenil, Jean-Paul Delahaye, and Nicolas Gauvrit.
\newblock Calculating {Kolmogorov} complexity from the output frequency
  distributions of small {Turing} machines.
\newblock {\em PloS one}, 9(5):e96223, 2014.

\bibitem{zenil2019coding}
Hector Zenil, Liliana Badillo, Santiago Hern{\'a}ndez-Orozco, and Francisco
  Hern{\'a}ndez-Quiroz.
\newblock Coding-theorem like behaviour and emergence of the universal
  distribution from resource-bounded algorithmic probability.
\newblock {\em International Journal of Parallel, Emergent and Distributed
  Systems}, 34(2):161--180, 2019.

\bibitem{legg2013approximation}
Shane Legg and Joel Veness.
\newblock An approximation of the universal intelligence measure.
\newblock In {\em Algorithmic Probability and Friends. Bayesian Prediction and
  Artificial Intelligence}, pages 236--249. Springer, 2013.

\bibitem{tang2015complexity}
Ling Tang, Huiling Lv, Fengmei Yang, and Lean Yu.
\newblock Complexity testing techniques for time series data: A comprehensive
  literature review.
\newblock {\em Chaos, Solitons \& Fractals}, 81:117--135, 2015.

\bibitem{torres2000relative}
ME~Torres and LG~Gamero.
\newblock Relative complexity changes in time series using information
  measures.
\newblock {\em Physica A: Statistical Mechanics and its Applications},
  286(3-4):457--473, 2000.

\bibitem{bialek2001complexity}
William Bialek, Ilya Nemenman, and Naftali Tishby.
\newblock Complexity through nonextensivity.
\newblock {\em Physica A: Statistical Mechanics and its Applications},
  302(1-4):89--99, 2001.

\bibitem{lloyd2001measures}
Seth Lloyd.
\newblock Measures of complexity: a nonexhaustive list.
\newblock {\em IEEE Control Systems Magazine}, 21(4):7--8, 2001.

\bibitem{turing1936computable}
Alan~Mathison Turing.
\newblock On computable numbers, with an application to the
  entscheidungsproblem.
\newblock {\em J. of Math}, 58(345-363):5, 1936.

\bibitem{grunwald2004shannon}
Peter Grunwald and Paul Vit{\'a}nyi.
\newblock {Shannon information and Kolmogorov complexity}.
\newblock {\em arXiv preprint cs/0410002}, 2004.

\bibitem{calude2002information}
C.S. Calude.
\newblock {\em Information and randomness: An algorithmic perspective}.
\newblock Springer, 2002.

\bibitem{gacs1988lecture}
P.~G{\'a}cs.
\newblock {\em Lecture notes on descriptional complexity and randomness}.
\newblock Boston University, Graduate School of Arts and Sciences, Computer
  Science Department, 1988.

\bibitem{buchanan2018natural}
Mark Buchanan.
\newblock A natural bias for simplicity.
\newblock {\em Nature Physics}, 14:1154, 2018.

\bibitem{alaskandarani2022low}
Mohamed Alaskandarani and Kamaludin Dingle.
\newblock Low complexity, low probability patterns and consequences for
  algorithmic probability applications.
\newblock {\em arXiv preprint arXiv:2207.12251}, 2022.

\bibitem{lempel1976complexity}
A.~Lempel and J.~Ziv.
\newblock On the complexity of finite sequences.
\newblock {\em Information Theory, IEEE Transactions on}, 22(1):75--81, 1976.

\bibitem{ziv1977universal}
Jacob Ziv and Abraham Lempel.
\newblock A universal algorithm for sequential data compression.
\newblock {\em IEEE Transactions on information theory}, 23(3):337--343, 1977.

\bibitem{willbrand2005identifying}
K.~Willbrand, F.~Radvanyi, J.P. Nadal, J.P. Thiery, and T.M.A. Fink.
\newblock Identifying genes from up--down properties of microarray expression
  series.
\newblock {\em Bioinformatics}, 21(20):3859--3864, 2005.

\bibitem{vitanyi2020incomputable}
Paul Vit{\'a}nyi.
\newblock How incomputable is kolmogorov complexity?
\newblock {\em Entropy}, 22(4):408, 2020.

\bibitem{cover1974universal}
TM~Cover.
\newblock Universal gambling schemes and the complexity measures of kolmogorov
  and chaitin. rep. no. 12, statistics dep, 1974.

\bibitem{langdon1983note}
GLEN Langdon.
\newblock A note on the ziv-lempel model for compressing individual sequences
  (corresp.).
\newblock {\em IEEE Transactions on Information Theory}, 29(2):284--287, 1983.

\bibitem{rissanen1984universal}
Jorma Rissanen.
\newblock Universal coding, information, prediction, and estimation.
\newblock {\em IEEE Transactions on Information theory}, 30(4):629--636, 1984.

\bibitem{feder1992universal}
Meir Feder, Neri Merhav, and Michael Gutman.
\newblock Universal prediction of individual sequences.
\newblock {\em IEEE transactions on Information Theory}, 38(4):1258--1270,
  1992.

\bibitem{cover2006elements}
TM~Cover and J.A. Thomas.
\newblock {\em Elements of information theory}.
\newblock John Wiley and Sons, 2006.

\bibitem{feder1991gambling}
Meir Feder.
\newblock Gambling using a finite state machine.
\newblock {\em IEEE Transactions on Information Theory}, 37(5):1459--1465,
  1991.

\bibitem{merhav2019universal}
Neri Merhav and Asaf Cohen.
\newblock Universal randomized guessing with application to asynchronous
  decentralized brute--force attacks.
\newblock {\em IEEE Transactions on Information Theory}, 66(1):114--129, 2019.

\bibitem{merhav1998universal}
Neri Merhav and Meir Feder.
\newblock Universal prediction.
\newblock {\em IEEE Transactions on Information Theory}, 44(6):2124--2147,
  1998.

\bibitem{plotnik1992upper}
Eli Plotnik, Marcelo~J Weinberger, and Jacob Ziv.
\newblock Upper bounds on the probability of sequences emitted by finite-state
  sources and on the redundancy of the lempel-ziv algorithm.
\newblock {\em IEEE transactions on information theory}, 38(1):66--72, 1992.

\bibitem{ryabko2016compression}
Boris Ryabko, Jaakko Astola, and Mikhail Malyutov.
\newblock {\em Compression-based methods of statistical analysis and prediction
  of time series}.
\newblock Springer, 2016.

\bibitem{ryabko2009compression}
Boris Ryabko.
\newblock Compression-based methods for nonparametric prediction and estimation
  of some characteristics of time series.
\newblock {\em IEEE Transactions on Information Theory}, 55(9):4309--4315,
  2009.

\bibitem{zenil2011algorithmic}
H.~Zenil and J.P. Delahaye.
\newblock An algorithmic information theoretic approach to the behaviour of
  financial markets.
\newblock {\em Journal of Economic Surveys}, 25(3):431--463, 2011.

\bibitem{fink20071}
T.M.A. Fink, K.~Willbrand, and F.C.S. Brown.
\newblock 1-d random landscapes and non-random data series.
\newblock {\em EPL (Europhysics Letters)}, 79(3):38006, 2007.

\bibitem{rissanen1978modeling}
Jorma Rissanen.
\newblock Modeling by shortest data description.
\newblock {\em Automatica}, 14(5):465--471, 1978.

\bibitem{grunwald2000model}
Peter Gr{\"u}nwald.
\newblock Model selection based on minimum description length.
\newblock {\em Journal of mathematical psychology}, 44(1):133--152, 2000.

\bibitem{bach2022information}
Francis Bach.
\newblock Information theory with kernel methods, 2022.

\bibitem{hamzi2021learning}
Boumediene Hamzi and Houman Owhadi.
\newblock Learning dynamical systems from data: A simple cross-validation
  perspective, part i: Parametric kernel flows.
\newblock {\em Physica D: Nonlinear Phenomena}, 421:132817, 2021.

\bibitem{schmidhuber1995discovering}
J{\"u}rgen Schmidhuber.
\newblock Discovering solutions with low kolmogorov complexity and high
  generalization capability.
\newblock In {\em Machine Learning Proceedings 1995}, pages 488--496. Elsevier,
  1995.

\bibitem{schmidhuber1997shifting}
J{\"u}rgen Schmidhuber, Jieyu Zhao, and Marco Wiering.
\newblock Shifting inductive bias with success-story algorithm, adaptive levin
  search, and incremental self-improvement.
\newblock {\em Machine Learning}, 28(1):105--130, 1997.

\bibitem{valle2018deep}
Guillermo Valle-P{\'e}rez, Chico~Q Camargo, and Ard~A Louis.
\newblock Deep learning generalizes because the parameter-function map is
  biased towards simple functions.
\newblock {\em arXiv preprint arXiv:1805.08522}, 2018.

\bibitem{zenil2019causal}
Hector Zenil, Narsis~A Kiani, Allan~A Zea, and Jesper Tegn{\'e}r.
\newblock Causal deconvolution by algorithmic generative models.
\newblock {\em Nature Machine Intelligence}, 1(1):58--66, 2019.

\bibitem{hernandez2020algorithmic}
Santiago Hern{\'a}ndez-Orozco, Hector Zenil, J{\"u}rgen Riedel, Adam Uccello,
  Narsis~A Kiani, and Jesper Tegn{\'e}r.
\newblock Algorithmic probability-guided machine learning on non-differentiable
  spaces.
\newblock {\em Frontiers in artificial intelligence}, 3, 2020.

\bibitem{hutter2007universal}
Marcus Hutter.
\newblock On universal prediction and bayesian confirmation.
\newblock {\em Theoretical Computer Science}, 384(1):33--48, 2007.

\bibitem{hutter2004universal}
Marcus Hutter.
\newblock {\em Universal artificial intelligence: Sequential decisions based on
  algorithmic probability}.
\newblock Springer Science \& Business Media, 2004.

\end{thebibliography}

\appendix
\newpage
\onecolumngrid

\section{The effect of changing $n$}

\begin{figure*}[htp]
\begin{center}
\subfigure[]{\label{fig:edge-a}\includegraphics[height=6.5cm,width=6.5cm]{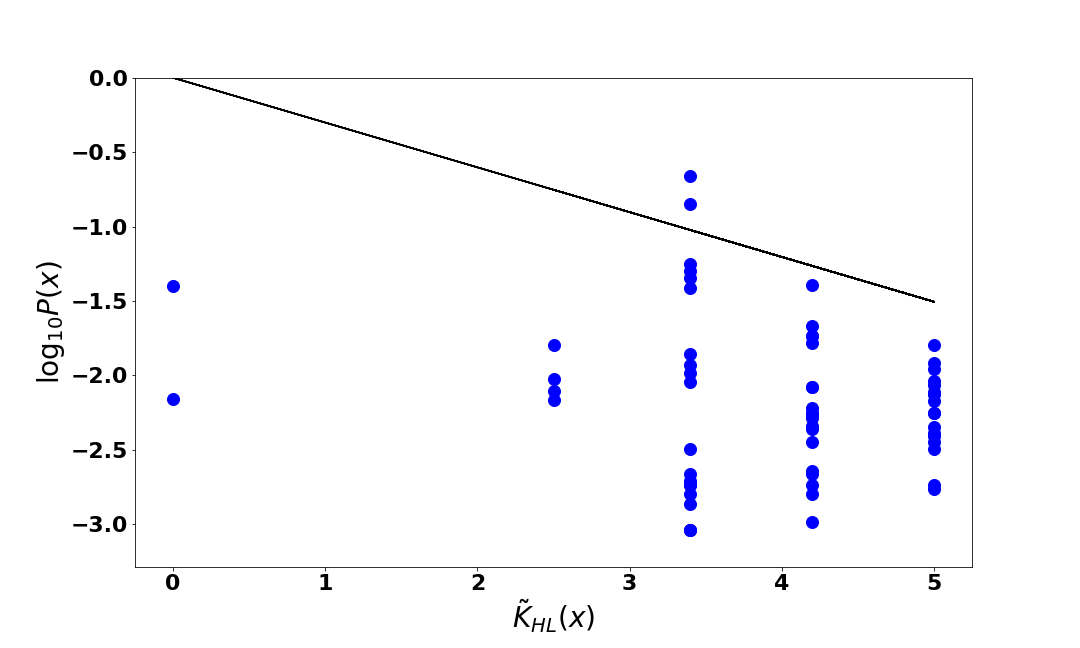}}
\subfigure[]{\label{fig:edge-a}\includegraphics[height=6.5cm,width=6.5cm]{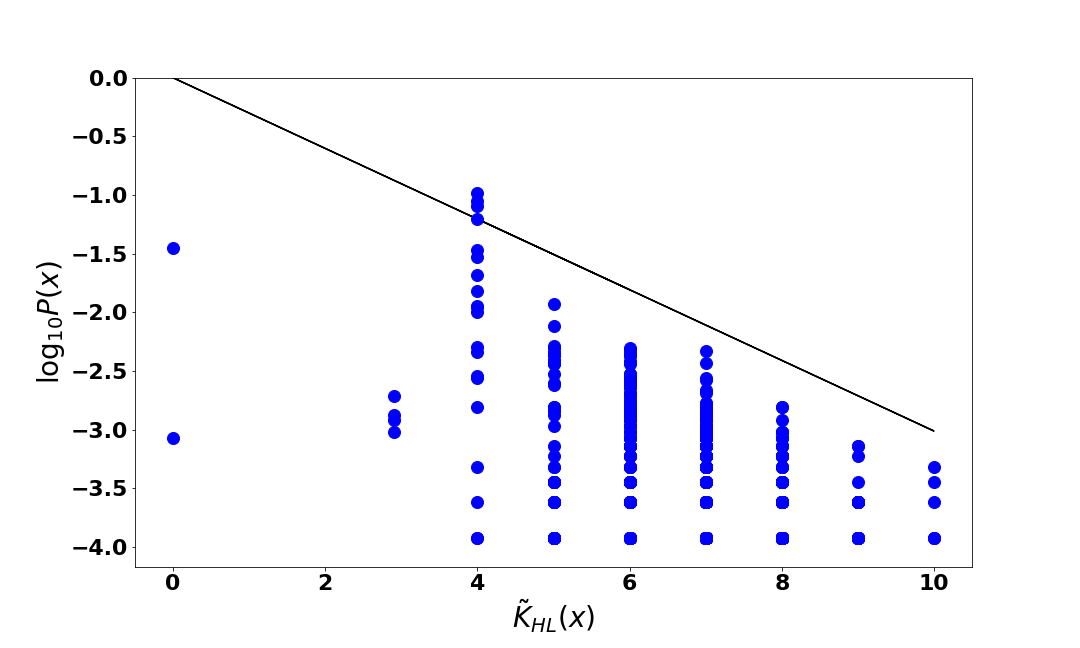}}
\subfigure[]{\label{fig:edge-a}\includegraphics[height=6.5cm,width=6.5cm]{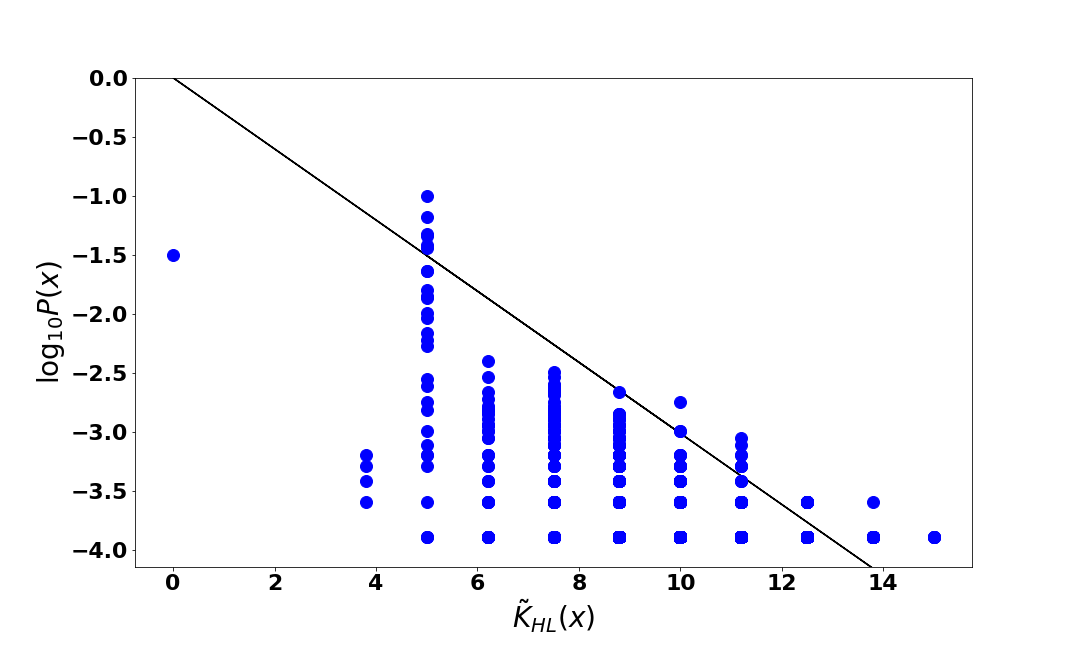}}
\subfigure[]{\label{fig:edge-a}\includegraphics[height=6.5cm,width=6.5cm]{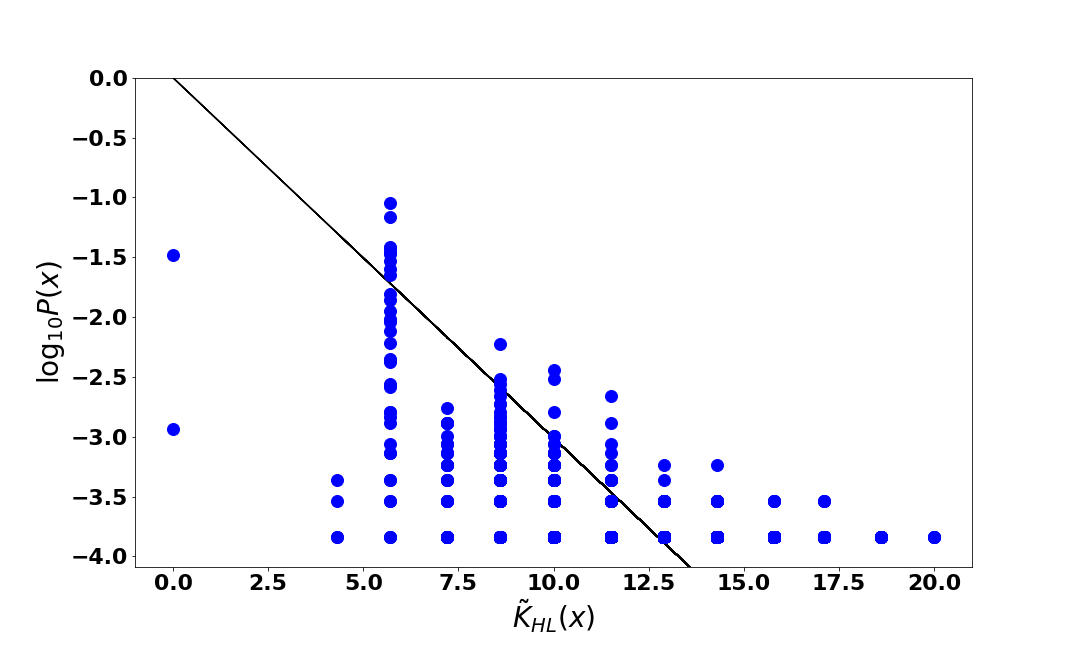}}

\end{center}
\caption{World development indicators data with HL method, for different lengths. (a) With $n=5$, SB is not observed. (b) With $n=10$ SB is observed and also in (c) with $n=15$. When $n$ gets too large for the data set as in (d) $n=20$, SB is less clear and more noise is apparent in the tail of the distribution. }
\label{fig:different_n}
\end{figure*}

In this section we illustrate the resulting patterns when for a given data set, different values of series length $n$ are used. Additionally, as another sector example, here we use data for world development indicators, again from the World Bank Open Data. Figure \ref{fig:different_n} shows complexity-probability plots for $n=5,10,15$ and 20 bit series. There are about 8,700 series in this data set. We focus on just the HL method in this section. With very small $n=5$, the plot is noisy with little clear relation between $P(x)$ and $\tilde{K}_{HL}(x)$. With $n=10$ and $n=15$, clear SB is observed in this data set. Moving up to $n=20$ SB is still observed, but because there are  exponentially more possible patterns, there are too few series data in this data set, leading to large errors in frequency estimates (hence $P(x)$). This leads to SB becoming a little less clear, especially in the tail at higher complexity values. Many of those high complexity patterns have over exaggerated frequencies due to low or single count frequencies per pattern.

\end{document}